\RequirePackage{etex}
\documentclass{aa}     
\usepackage{float}
\usepackage{graphicx}   
\usepackage{txfonts}    
\usepackage{hyperref}
\hypersetup{
    colorlinks=true,
    linkcolor=blue,
    filecolor=magenta,      
    urlcolor=blue,
    citecolor=blue,   
    }
\usepackage{natbib}
\usepackage{algpseudocode}
\usepackage{amssymb} 
\usepackage{amsmath}
\usepackage{breqn}
\usepackage[toc, page]{appendix}
\bibpunct{(}{)}{;}{a}{}{,} 

\title{DESTINY: a new binding-energy-resolved astrochemical framework}

\subtitle{Self-Consistent Competitiveness using Branched Absorbing Markov Chains}

\author{Maria Groyne\inst{1} \and Cedric Baijot\inst{1} \and Michaël De Becker\inst{1}} 

\institute{Space Sciences, Technologies and Astrophysics Research (STAR) Institute, University of Liège, Quartier Agora, 19c, Allée du 6 Août, B5c, 4000, Sart Tilman, Belgium}

\date{Received June 19, 2026; accepted July 17, 2026}

\abstract
{
Under cryogenic interstellar conditions, the amorphous structure of interstellar ice results in binding-energy distributions (BEDs) per species. However, only few studies attempted their inclusion in astrochemical models. 
}
{
This paper introduces DESTINY, a deterministic astrochemical framework designed to incorporate BEDs while self-consistently accounting for the competition among activated surface processes. The framework is currently constrained to a monolayer.
}
{
Surface processes initiated by surface species are reformulated using a trial-frequency-capped formalism represented through branched absorbing Markov chains. The ordinary differential equations (ODE) system is redefined based on normalized effective probabilities.  
}
{ 
Preliminary results based on a reduced surface network are discussed. To isolate the effects of the probabilistic reformulation from those of BED discretizations, DESTINY is benchmarked against Nautilus, a single-BE rate-equation based open source code. In the single-BE limit, DESTINY reproduces the behavior of Nautilus for most species. The largest deviations are obtained for CH$_{x\in [2,4]}$; these are traced to a different treatment of the H$_2$ encounter effect, impacting both H$_2$ surface exploration and desorption efficiencies within the DESTINY framework. Introducing BEDs redistributes species among adsorption sites of different depths, altering the balance between diffusion, desorption, and reactions. Significant effects are found for H, H$_2$, NH$_x$, NO, CH$_x$, CO and H$_x$CO. 
}
{
Preliminary results showed that the self-consistent treatment of the H$_2$ encounter effect coupled with the explicit treatment of BEDs can substantially modify grain-surface chemistry. Further framework extensions are expected in the near future. 
}

\keywords{Astrochemistry -- ISM: abundances -- ISM: molecules -- ISM: clouds -- Methods: numerical}

\begin{document}
\titlerunning{DESTINY: I.}
\maketitle
\nolinenumbers 

\section{Introduction}
\label{sec:introduction}

The chemical diversity of the interstellar medium (ISM) emerges from the dynamical interplay between gas-phase and dust grain chemistry. Dust grains are key actors, as they act both as passive third bodies dissipating reaction energy excess and as active catalytic surfaces, while maintaining reactants in proximity \citep{2024ARA&A..62..243C}. It opens chemical pathways that cannot operate efficiently in the gas phase, essential  to account for the observed abundances of fundamental species such as H$_2$ and H$_2$O \citep{1963ApJ...138..393G, 1971ApJ...163..155H, 1972ApJ...174..321W, 1982A&A...114..245T}, while facilitating the build-up of molecular complexity in the coldest part of the ISM \citep{2013ApJ...765...60G, 2025A&A...702A.236M, 2025ApJS..277....8L, 2025A&A...701A.131N}. In addition, gas-grain interactions through adsorption and desorption dynamically couple the solid and gas phases, strongly shaping the chemical composition of the ISM \citep{2020ARA&A..58..727J, 2022ApJ...926..171F, 2023A&A...670A.141T, 2025A&A...694A.263T, 2025A&A...698A.278T, 2026A&A...705A.185B}. To investigate and predict the chemical evolution of astrophysical environments, astrochemical kinetic models have been developed. To this end, the rate equation (RE) approach is widely used \citep{1992ApJS...82..167H, 1993MNRAS.261...83H,
1993MNRAS.263..589H, 2016MNRAS.459.3756R, 2017AJ....154...38H, 2021A&A...652A..63W, 2024A&A...689A..63W, Angele_Chem_Des, 2025A&A...699A.332M}. This framework relies on deterministic mean-field kinetics that neglect the stochasticity and microscopic spatial structure of dust grains. Stochastic approaches, such as kinetic Monte Carlo (kMC) methods \citep{2013ChemRev.113.12.8840, 2018ApJ...869..165L}, overcome several limitations of traditional RE-based models by explicitly simulating individual chemical events and naturally accounting for fluctuations arising from the discrete nature of grain chemistry \citep{2009A&A...508..275C, 2009ApJ...691.1459V, 2009ApJ...700L..43G}. However, the inherent event-by-event treatment comes at a significant computational cost, which becomes prohibitive for large chemical networks and long timescales, making deterministic RE-based models the method of choice for many astrochemical applications. A recent benchmark study has nevertheless highlighted the sensitivity of grain-surface chemistry predictions to the adopted formalism, with discrepancies arising both between deterministic and stochastic approaches and among deterministic formulations themselves \citep{2025A&A...695A.247J}.

Yet, the complexity of dust grain chemistry limits the ability of deterministic RE-based approaches to faithfully reproduce the evolution of the chemical reservoir adsorbed on dust grains. One of the main simplifying assumptions is the use of a single binding energy (BE) value per species to characterize its dynamical journey across the ice, both in terms of mobility and residence time \citep{2017SSRv..212....1C}. Nevertheless, under the cryogenic conditions prevailing in dark molecular clouds, the icy mantle covering dust grains is known to be of an amorphous, water-rich nature \citep{1989ApJ...344..413S, 2013ChRv..113.8783H,2015ARA&A..53..541B, 2024A&A...688A..29S}. This amorphous character leads to a rich diversity of possible substrate-to-adsorbate binding configurations, resulting in a BE distribution (BED) per ice constituent \citep{Bovolenta, Ferrero, Duflot, Tinacci, 2025A&A...698A.284G}. While strongly bound configurations favor kinetic trapping effects, shallower sites increase desorption efficiency as well as mobility, which is essential to activate the Langmuir-Hinshelwood (LH) reactive path \citep{2012ApJ...757..185H, 2024ApJ...974..115F}.
Recent efforts have been made towards inferring an evolving set of BEDs for relevant interstellar species on amorphous icy substrates \citep{Karssemeijer, Song_Kastner, Bovolenta, Ferrero, Molpeceres_Kastner, Duflot, 2022ESC.....6..597M, Tinacci, 2025A&A...698A.284G, 2025ApJ...993..184K, 2026LSSR...49...77R}. This constitutes the necessary prerequisite for moving beyond the traditional single BE approximation, the first step toward a more realistic description of solid-phase astrochemistry. Although only a few studies have attempted to incorporate BEDs into astrochemical  models, a consistent conclusion emerges: accounting for BE heterogeneity significantly alters grain-surface chemistry and its coupling with the gas phase \citep{2011A&A...529A.151C, 2019ApJ...882..172F,2020A&A...643A.155G, 2024ApJ...974..115F, 2026ApJ..1001...62F}. For instance, changes in H$_2$ formation efficiencies were already demonstrated by \citet{2011A&A...529A.151C} using a continuous-time random-walk Monte Carlo method \citep{2005A&A...434..599C} with two BEs for H. \cite{2020A&A...643A.155G} expanded the analyses to more species within a newly proposed deterministic framework, using up to 51 bins. Yet, this treatment was not fully self-consistent, as the diffusion only served the LH mechanism without considering inter-bin redistribution. The latter has been formalized in \cite{2024ApJ...974..115F}. While the use of bins in \cite{ 2020A&A...643A.155G} increased the number of ODEs to be solved and, inherently, the computational cost, the deterministic framework proposed in \cite{2024ApJ...974..115F} introduced a probability density function (PDF) to approximate the occupation of binding sites (BS) without additional ODEs.

However, some limitations remain \citep{2024ApJ...974..115F}. For instance, the PDF is assumed to adjust instantaneously to time-varying physical conditions. Another constraint is the restriction to one representative of each species per BS, although up to $n$ distinct species may coexist on the same site. The resulting vacancy factors associated with diffusive species (Eq. 20-21, 26, 28-29 in \cite{2024ApJ...974..115F}) have consequences that extend beyond the constraint of site availability; they inherently affect surface exploration efficiency, especially at high coverage for the diffusing species. Considering a surface with a non-negligible coverage fraction from species $j$ searching for a $k$ constituent, the vacancy factor for a diffusive $j$ artificially  reduces its effective mobility and, consequently, its surface exploration efficiency and mean exploration path for locating a reaction partner. 

Another open challenge persists in the self-consistent treatment of surface process competitiveness. For instance, although reaction–diffusion competition is accounted for when evaluating activated reaction rates (e.g., Eqs. 28-29 in \cite{2024ApJ...974..115F}, based on the single-BE-based prescriptions in \cite{2011ApJ...735...15G}), its feedback onto the diffusive flux describing the redistribution of occupation over the BE landscape is not treated, as the inter-bin redistribution is decoupled from the competing reactive channel. A concrete example concerns the H$_2$ encounter desorption mechanism discussed in \citet{Hincelin2015}. Physically, this mechanism implies a strong coupling between diffusion and desorption: upon encountering another H$_2$ molecule, the effective BE of H$_2$ is substantially reduced, which should increase its desorption probability as well as its mobility and associated surface exploration efficiency. This mechanism will consequently be called H$_2$ encounter effect rather than H$_2$ desorption encounter mechanism for the rest of the paper. Yet, in \citet{2024ApJ...974..115F}, diffusion and desorption are treated as independent fluxes, such that the enhanced mobility associated with encounter-induced BE lowering is not naturally captured. As proposed in \citet{Hincelin2015}, the H$_2$ encounter effect can be introduced as an additional surface reaction channel H$_2$ + H$_2$, allowing one of them to desorb based on the diffusion–desorption competition factor, computed with the lowered BE. However, while the framework in \citet{2024ApJ...974..115F} accounts for inter-bin diffusive redistribution, it does not couple the non-desorbing H$_2$ encounter flux to the inter-bin redistribution channel, thereby neglecting the associated enhancement of H$_2$ mobility.

While the pioneering works proposed in \cite{2020A&A...643A.155G, 2024ApJ...974..115F} constitute a major step forward towards BE-resolved deterministic astrochemical modelling, important open challenges persist, especially in the treatment of diffusion-driven processes and their interplay with desorption and reactivity across heterogeneous BE landscapes. Altogether, these challenges motivate the further development of the emerging generation of BE-resolved deterministic astrochemical frameworks. To this end, this paper introduces DESTINY (Discretized binding Energy-based grain Site coverage for Trial frequency capped Integrated kinetics with Normalized Yields), designed to self-consistently treat the interdependence arising from the mutual competitiveness of diffusion, desorption, and reaction processes, enforcing a proper redistribution of chemical fluxes across the BE landscape. Surface processes are represented through a probabilistic adsorbing Markov chain representation, naturally accounting for process competition and flux branching. While DESTINY assumes that each BS can host at most one species at the ODE level, a transient subspace permits temporary double BS occupancy. Idle states are introduced to ensure probability conservation. The stationary distribution of all possible outcomes from the current state towards any transient or absorbing states enables the branching of chemical fluxes into BE-discretised normalised effective rates, which are subsequently injected into the ODE system right-hand side (RHS).

The mathematical framework is presented in Sect. \ref{sec: framework}. Code-specific numerical treatments are commented in Sect. \ref{Sect: Numerical_code}. In Sect. \ref{sec: result}, the first DESTINY results are discussed, based on a reduced chemical network with static physical parameters. The discussion is primarily focused onto (i) a comparative study with a single-BE RE-based open source code, and (ii) the quantification of the relative impact of the BEDs onto the simulated grain chemistry. Conclusions and outlooks are presented in Sect. \ref{sec: ccl_perspect}.

\section{General DESTINY Framework}\label{sec: framework}

Provided that the BEDs are sampled over a sufficiently large number of BSs, they can be assumed to be representative of an entire grain surface. In this context, the normalized BED of species $i$ can be interpreted as the PDF of BSs with a BE $E_i$, denoted $g_i(E_i)$, naturally satisfying Eq. \ref{eq: g_i_E_i}.

\begin{equation}
\label{eq: g_i_E_i}
    \int_{0}^{\infty} g_i (E_i) \, dE_i = 1
\end{equation}

The BE-resolved coverage $\theta_i(E_i)$ denotes the fraction of BSs $E_i$ filled by species $i$. 
Averaging over the BED yields the surface coverage of the species $i$, as given in Eq. \ref{eq: theta_i_E_i} \citep{2024ApJ...974..115F}.

\begin{equation}
\label{eq: theta_i_E_i}
    \int_{0}^{\infty} \theta_i(E_i) g_i(E_i) \, dE_i = \Theta_i = \frac{<N_i>}{N_{site}}
\end{equation}

\noindent where $<N_i>$ is the average number of species $i$ on a single dust grain, and $N_{site}$ is the number of sites per grain surface. The surface is therefore described statistically as a collection of $N_{site}$ BSs without explicit spatial resolution. 

We also introduce $\Theta$, the total fractional coverage of a monolayer summed over all surface species. The definition of $\Theta$ is based on the DESTINY working hypothesis that each BS can be filled at most by one species at the ODE solver level. Within the monolayer approximation, $\Theta$ is therefore constrained by Eq. \ref{eq: theta_cdt}. 

\begin{equation}
\label{eq: theta_cdt}
    \sum_i \Theta_i = \Theta \leq 1
\end{equation}

The temporal evolution of the solid-phase system is computed from a set of coupled ODEs that describe the time evolution of each $\theta_i(E_i)$, as explained in the following sub-sections. 

\subsection{Trial Frequency-Capped Monolayer Formalism}
\label{sec: tfc-formalism}

In order to properly describe the dynamics of surface chemistry, the frequency at which a given species attempts to initiate any activated process has to be quantified. In the context of astrochemical kinetics, this characteristic attempt rate is generally identified with the trial frequency, $\nu_i (E_i)$. Following \citet{1987ASSL..134..397T} and \citet{1992ApJS...82..167H}, the trial frequency is approximated as the vibrational frequency of an adsorbate oscillating within its surface potential well. Under the harmonic approximation of the adsorption potential, it is given by Eq. \ref{eq: trial_frequency}

    \begin{equation}
        \nu_i(E_i) = \left(\frac{2\,n_{\rm site}\,E_i}{\pi^2\,m_i}\right)^{1/2},
        \label{eq: trial_frequency}
    \end{equation}  

\noindent where $n_{\rm site}$ is the surface site density, and $m_i$ is the species mass \citep{1987ASSL..134..397T, 1992ApJS...82..167H}. It therefore represents a capping agent imposing an upper bound on solid-phase process initiation, defining the intrinsic event-initiation clock of species $i$. Consequently, capping-relevant events must mutually compete for this finite budget, such that their collective initiation rate cannot exceed $\nu_i$. This competition naturally applies to all processes whose kinetics are governed by the crossing of an energetic barrier separating two states.

It is worth mentioning that this limitation only applies to processes actively initiated by a grain species $i$, namely non-reactive and reactive diffusion, as well as thermal desorption. In contrast, adsorption and Eley--Rideal (ER) reactions are triggered by incoming gas-phase species and therefore do not draw from the dynamical budget of the solid-phase species. The same applies to UV- and CR-photon-induced desorption. Their competition with surface-initiated events is instead naturally captured through the coupling of the ODE system and the associated conservation constraints, excluding them from the trial-frequency capping constraint.

\subsubsection{Branched Markov Chain level description}

\begin{figure*}[t]
    \centering
    \includegraphics[width=0.83  \linewidth]{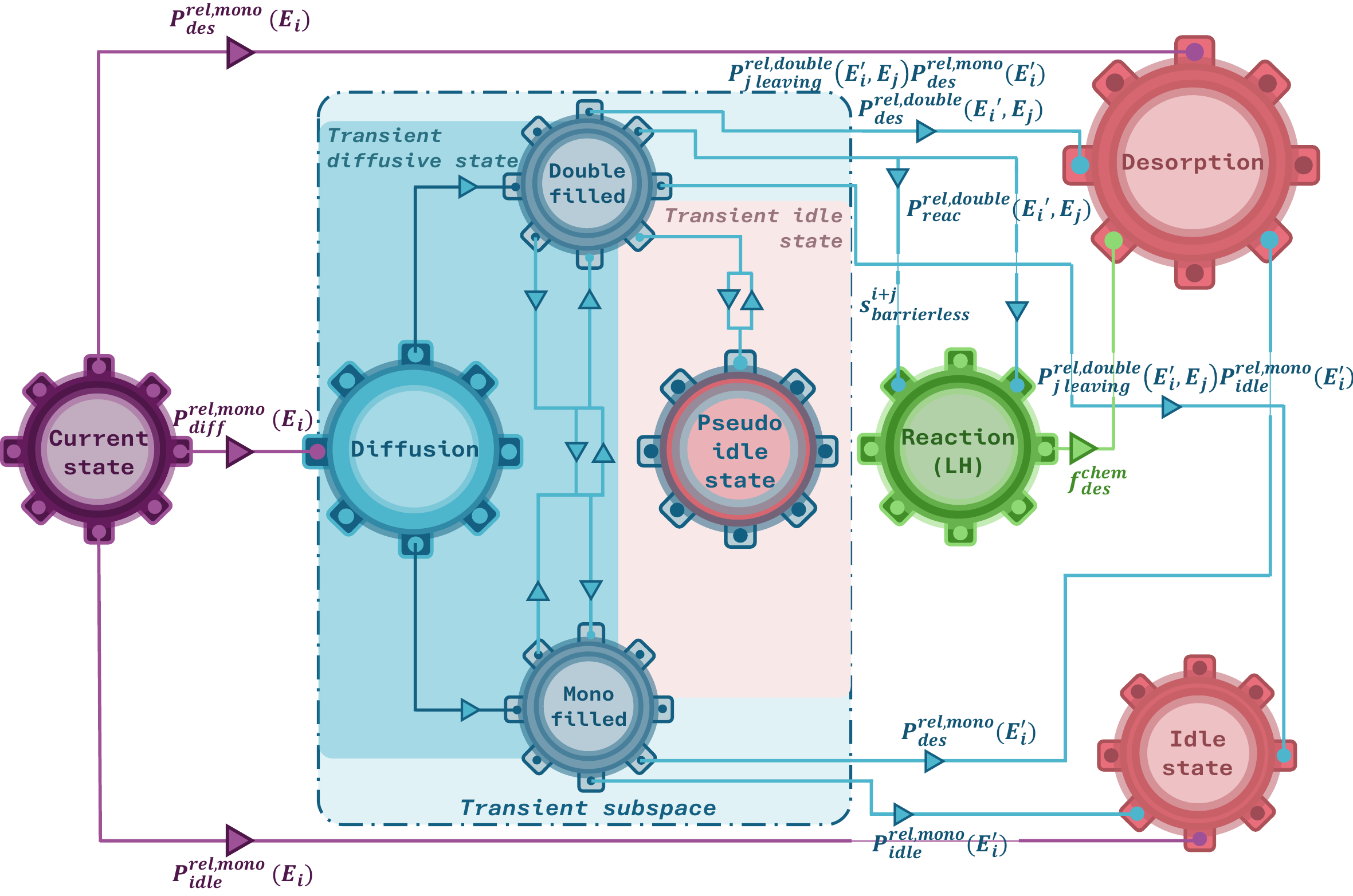}
    \caption{Branched absorbing Markov chain representation of a trial frequency-capped (TFC) monolayer system evolution. Red and green gears represent respectively absorbing and quasi-absorbing states, while the blue box concerns the transient subspace and its diffusive and pseudo-idle sub-regions. See the text for the full explanation. }
    \label{fig:Markov Chain}
\end{figure*}

As previously stated, the proposed framework aims to self-consistently treat inter-process competition within BE-resolved surface populations. To this end, the evolution of a trial frequency-capped (TFC) monolayer system over a time step is represented as a succession of attempts for competing event initiation, leading to distinct possible outcomes. The competition between all accessible processes should be explicitly evaluated at each attempt, ensuring a consistent treatment of process competitiveness throughout the system evolution. Under the assumption that the event sequence satisfies the Markov property, such that each event depends solely on the instantaneous state, this TFC system representation naturally maps onto a branched Markov chain with absorbing states, as schematically depicted in Fig. \ref{fig:Markov Chain}. An absorbing state defines here any terminal state from which the considered species cannot initiate any further surface processes within the current sequence. This includes desorption and idle states. Surface reactions are treated as quasi-absorbing states: they behave as absorbing states within the TFC Markov-chain formalism but may subsequently be followed by a desorption event through the coupled ODE framework (Sect. \ref{Sect: ODE system formulation}). Their quasi-absorbing nature reflects this conditional irreversibility: in the absence of post-reaction desorption, the reaction product remains adsorbed, and the sequence terminates. In Fig. \ref{fig:Markov Chain}, absorbing and quasi-absorbing states are shown in red and green, respectively.

Between the initial state (initial state of the chain, i.e. current state for the ODE solver, mauve node in Fig. \ref{fig:Markov Chain}) and the absorbing outcomes, species $i$ may enter a transient subspace (blue box in Fig. \ref{fig:Markov Chain}) composed of diffusive and pseudo-idle states. The dynamics of this non-absorbing subregion of the Markov chain unfolds through a sequence of internal diffusive and pseudo-idle states prior to termination. We define a pseudo-idle state as a transient idle configuration on a double-filled site, which is considered an idle state within the transient subspace but is non-terminal with respect to the Markov chain. In contrast, the absorbing idle state represents the terminal outcome of a transient sequence not ended by a reaction or desorption. To remain consistent with the ODE-level single-occupancy constraint of Eq. \ref{eq: theta_cdt}, absorbing states are restricted to mono-filled site outcomes, whereas double-filled configurations are confined to the transient subspace. Consequently, chains are both initiated and terminated on singly occupied sites, while double-filled states contribute exclusively through transient dynamics.

More specifically, from the current state node, the flux associated with species $i$ at a site of binding energy $E_i$ is partitioned into three main branches

    \begin{itemize}
        \item [I.] Direct thermal desorption from the current site, with direct access to the absorbing desorption state;

        \item [II.] Idle state on the current site, leading to a direct transition to the absorbing idle state from the starting mono-filled site; 

        \item[III.] Entrance within the transient substate. The unique entry point to the transient state is a diffusion event from the current mono-filled site. This acts as a gateway to the transient diffusive substate, from which all pathways of the transient dynamics (light blue arrows) originate.

        Once entered, the transient state is composed of two coupled substates: a transient idle substate on double-filled sites and a transient diffusive substate (i.e., site-to-site diffusion on vacant or filled site, leading to mono- or double-filled site). The transient state persists until sub-chain termination towards adsorbing states, occuring through three distinct pathways. 
        
        First, it may follow a non-reactive diffusive pathway without any subsequent event, ending in the absorbing idle state. This occurs either after landing on an empty site (mono-filled configuration), or after landing on a site occupied by species $j$, leading to a double-filled site, provided that species $j$ leaves the site through diffusion or desorption before any action of $i$. The latter case restores a mono-filled configuration, consistent with the ODE-level mono-occupation constraint. Alternatively, the sub-chain may proceed through a non-reactive diffusive pathway with a downstream desorption from either a mono-filled or a double-filled site.

        The third transient sub-chain termination pathway corresponds to reactive diffusion after landing on an occupied site, leading to a double-filled configuration (LH-like mechanism). The presence of an activation barrier introduces two distinct LH regimes, requiring separate treatments within the TFC formalism. In the general case, the reactive channel constitutes a competitive branch alongside the other TFC-relevant processes. In the barrierless limit, however, reaction occurs immediately upon formation of the double-filled configuration (encounter), bypassing the competitive branch. Outside the TFC subsystem, the resulting product may subsequently undergo chemical desorption (Sect. \ref{Sect: ODE system formulation}).

    \end{itemize}

From a mathematical perspective, the absolute probabilities associated with each trial-initiated process should first be defined. These quantities describe the intrinsic probability of the individual processes when considered independently, prior introducing competition with other TFC-relevant events. 

The absolute probability of thermally-promoted desorption from a species $i$, $P_{des} (E_i)$, is evaluated from the Boltzmann factor associated with the BE $E_i$, $e^{- E_i / T}$, with $T$ being defined by the thermal state of the grain. Within the present framework, two contributions are considered: (i) pure thermal desorption evaluated at the quiescent dust temperature $T_{dust}^{quiescent}$ ($P_{des}^{T_{dust}^{quiescent}}$, Eq. \ref{eq: Pdes_T}), and (ii) CR-heating-assisted desorption evaluated at the peak temperature $T_{peak}^{CR-heating}$ reached during transient cosmic-ray (CR) heating events ($P_{des}^{CR-heating} (E_i)$, Eq. \ref{eq: Pdes_th_CR}). $T_{peak}^{CR-heating}$ is generally set to 70 K \citep{1993MNRAS.261...83H}.

\begin{equation} \label{eq: Pdes_T}
    P_{des}^{T_{dust}^{quiescent}} (E_i) = e^{- E_i / T_{dust}^{quiescent}}
\end{equation}

\noindent with $E_i$ and $T_{dust}$ in K. Hereafter, all energy barriers are in K.

\begin{align} \label{eq: Pdes_th_CR}
    P_{des}^{CR-heating} (E_i) &= f_{peak}^{CR-heating} e^{- E_i / T_{peak}^{CR-heating}} \\ 
    \notag
    f_{peak}^{CR-heating} & = min\bigg(1, \ \zeta_{Fe+}^{\rm Grain} \Delta t_{peak}^{CR-heating} \bigg(\frac{\zeta_{CR}}{1.3 \times 10^{-17}}\bigg)\bigg)
\end{align}

\noindent where $f_{\rm peak}^{\rm CR-heating}$ is the fraction of time during which a grain resides near the transient CR-heating temperature peak \citep{1993MNRAS.261...83H}. Within the probabilistic interpretation adopted here, it is equivalent to the probability that a surface species experiences a CR-heating episode, assuming a thermal equilibration timescale between the grain and the adsorbate shorter than the peak duration. Following \citet{1993MNRAS.261...83H}, $f_{\rm peak}^{\rm CR-heating}$ is estimated as the product of the cooling time after a heating event, $\Delta t_{\rm peak}^{\rm CR-heating}$, driven by desorption from volatiles, and the effective frequency $\zeta_{\rm Fe+}^{\rm Grain}$ of whole-grain heating events induced by heavy CR nuclei (Fe$^+$). The latter is derived from the Fe$^+$ flux and the grain geometrical cross section, and corresponds to the inverse time interval between successive efficient heating events. $\zeta_{\rm CR}$ denotes the CR ionization rate.

The two thermal equilibrium states, characterized by $T_{\rm dust}^{quiescent}$ and $T_{\rm peak}^{\rm CR-heating}$, are treated as mutually exclusive. The corresponding desorption probabilities are therefore combined through the law of total probability. This follows from the assumption of thermal equilibrium between the adsorbate and the grain surface, together with the requirement that the characteristic thermal equilibrium timescale remains shorter than the duration of a CR-heating episode. Under these conditions, a species experiences either the equilibrium grain temperature $T_{\rm dust}$ or the transient peak temperature $T_{\rm peak}^{\rm CR-heating}$ at a given instant. The absolute desorption probability is therefore obtained from Eq. \ref{eq: Pdes}.

\begin{align}
    \label{eq: Pdes}\notag
    P_{des}(E_i)
    &=
    P_{T_i = T_{dust}}\,P_{{des}\,|\,T_{\rm dust}}(E_i)
    +
    P_{T_i = T_{\rm peak}^{\rm CR-heating}}\,
    P_{{des}\,|\,T_{\rm peak}^{\rm CR-heating}}(E_i)
    \\\notag
    &=
    \left(1-f_{\rm peak}^{\rm CR-heating}\right)
    e^{- E_i / T_{dust}}
    +
    f_{peak}^{CR-heating}\, e^{- E_i / T_{peak}^{CR-heating}}
    \\
    &=
    \left(1-f_{\rm peak}^{\rm CR-heating}\right)
    P_{des}^{T_{dust}^{quiescent}} (E_i)
    +
    P_{des}^{CR-heating}(E_i)
\end{align}

Analogous to the desorption case, the absolute diffusion probability, $P_{diff}(E_i)$, for species $i$ from a BS with BE $E_i$ towards any BS is constructed by distinguishing the two mutually exclusive thermal states of the grain: the quiescent equilibrium state at $T_{\rm dust}^{\rm quiescent}$ and the transient CR-heated state at $T_{\rm peak}^{\rm CR-heating}$. Thermally promoted hopping is conditioned by the thermal state of the grain and is therefore combined through the law of total probability. Quantum tunneling, by contrast, is treated as an independent diffusion pathway not affected by $T_{dust}$. The total diffusion probability is obtained by combining the thermally promoted hopping probability with the tunneling contribution as independent pathways. The absolute probability of thermally promoted hopping, $P_{diff}^{T}(E_i)$, is given by Eq. \ref{eq: Pdiff_thermal}.

\begin{align} \label{eq: Pdiff_thermal}
    P_{diff}^{T} (E_i) 
     & = \int e^{- \frac{E_{hop} (E_i  \rightarrow E_i')}{T}}  g_i(E_i') \, dE_i' 
\end{align}

\noindent The activation barrier for diffusion, $E_{hop}(E_i\rightarrow E_i')$, is given by Eq. \ref{eq: E_hop}, as used in \cite{2024ApJ...974..115F}. It depends on both the initial and final BEs $E_i$ and $E_i'$, based on the consideration in \cite{2017APJ.849.80.12} founded on microscopic reversibility.

\begin{equation}
\label{eq: E_hop}
    E_{hop} (E_i  \rightarrow E_i') = \chi \times min(E_i, E_i') + max(0, E_i - E_i')
\end{equation}
\noindent where $\chi$ is the hopping-to-binding energy ratio.
The absolute probability of a CR-heating-assisted diffusion of species $i$ from a site with BE $E_i$, $P_{diff}^{CR-heating}(E_i)$, is obtained by multiplying Eq.\,\ref{eq: Pdiff_thermal} evaluated at $T_{\rm peak}^{\rm CR-heating}$ by $f_{\rm peak}^{\rm CR-heating}$. 

The absolute quantum-tunelling-mediated diffusion probability, $P_{diff}^{tunn} (E_i)$, is expressed as in Eq. \ref{eq: Pdiff_tunn} under the approximation of rectangular activation barriers of thickness $a$, $P_{diff}^{tunn} (E_i)$. 

\begin{equation} \label{eq: Pdiff_tunn}
    P_{diff}^{tunn} (E_i) = \int e^{\left(-2a_{diff} \sqrt{\frac{2 \mu E_{hop} (E_i  \rightarrow E_i')}{\hbar^2}}\right)}  g_i(E_i') dE_i'
\end{equation}

\noindent with $\mu$, the reduced mass, and $a_{diff}$, the diffusion barrier width.

Combining the independent tunneling-assisted and thermally-activated contributions while considering the mutual exclusivity of the dust thermal state, the absolute probability of site-to-site diffusion, $P_{diff} (E_i)$, is given by Eq. \ref{eq: Pdiff}. 

\begin{align}\label{eq: Pdiff}
    P_{diff}(E_i)
    &=
    1-
    \left(
    1-P_{diff}^{thermal}(E_i)
    \right)
    \left(
    1-P_{diff}^{tun}(E_i)
    \right) \\\notag
    P_{diff}^{thermal}&(E_i)=
    \left(
    1-f_{\rm peak}^{\rm CR-heating}
    \right)
    P_{diff}^{T_{dust}^{quiescent}}(E_i) +
    P_{diff}^{CR-heating}(E_i)
\end{align}

The intrinsic probability of reaction upon encounter, $P_{cross}^{i, j}$, is constructed analogously to the diffusion probability. The mutually exclusive thermal states and related thermal crossing probabilities conditioned on the thermal state are combined through the law of total probability. In contrast, this combined thermally-promoted contribution and quantum tunneling probabilities are treated as independent pathways towards the same reactive outcome. The corresponding thermally activated, CR-heating-assisted, and quantum-tunneling-mediated probabilities are denoted $P_{T_{dust}}^{i,j}$, $P_{T_{peak}^{CR-heating}}^{i,j}$, and $P_{tunn}^{i,j}$, respectively. 

\begin{align}\label{eq: Pcross}
    &P_{cross}^{i,\, j}
    =
    1-
    \left(
    1-P_{thermal}^{i, j}
    \right)
    \left(
    1-P_{tunn}^{i, j}
    \right) \\
    \notag
    &P_{thermal}^{i, j}
    =
    \left(
    1-f_{\rm peak}^{\rm CR-heating}
    \right)
    P_{T_{dust}^{quiescent}}^{i, j} +
    P_{CR-heating}^{i, j}.
\end{align}

\noindent with a reaction barrier $E_A$ and an associated barrier width $a_{reac}$,

\begin{align}
    &P_{tunn}^{i, j} = e^{\left(-2a_{reac} \sqrt{\frac{2 \mu E_{A}}{\hbar^2}}\right)} \,\text{ \& }\,   P_{T_{dust}}^{i,j} =  e^{- \frac{E_A^{i, j}}{T_{dust}}} \notag \\
    &P_{CR-heating}^{i,j} = f_{\rm peak}^{\rm CR-heating} P_{T_{peak}^{CR-heating}}^{i,j} \notag
\end{align}

From these absolute probabilities, the absolute probability of TFC evolution of the system can be defined. As previously discussed, mono-filled and double-filled configurations must be treated separately. The treatment of the mutual exclusivity of thermal states and the independence of microscopic pathways within a given event type has already been embedded in the corresponding absolute probabilities. The absolute probabilities for desorption, diffusion, and reaction barrier crossing (Eqs. \ref{eq: Pdes}, \ref{eq: Pdiff}, and \ref{eq: Pcross}, respectively) can therefore be combined as independent evolution channels. For a species $i$ occupying a mono-filled site of binding energy $E_i$, the probability of evolving away from the mono-filled configuration through the accessible evolution channels, namely diffusion and desorption, is given by Eq. \ref{eq: Pevol_on_mono}.

\begin{align} \label{eq: Pevol_on_mono}
    P_{evol}^{mono} (E_i) &= 1 -(1 - P_{diff}(E_i)) (1 - P_{des} (E_i))
\end{align}

The relative diffusion and desorption probabilities, respectively $P_{diff}^{rel, mono}(E_i)$ and $P_{des}^{rel, mono}(E_i)$, are obtained by redistributing the evolution probability according to the relative weights of the diffusion and desorption channels. Equivalently, they correspond to the probabilities of diffusion and desorption conditioned on the occurrence of an evolution event, multiplied by the total evolution probability, as given in Eqs. \ref{eq: Pdiff_eff_from_mono} and \ref{eq: Pdes_eff_from_mono}.

\begin{align}
    \label{eq: Pdiff_eff_from_mono}
    &P_{diff}^{rel, mono} (E_i) = \frac{P_{diff} (E_i)}{P_{diff}(E_i) + P_{des} (E_i)} \times P_{evol}^{mono} (E_i) \\ 
    \label{eq: Pdes_eff_from_mono}
    &P_{des}^{rel, mono} (E_i) = \frac{P_{des} (E_i)}{P_{diff}(E_i) + P_{des} (E_i)} \times P_{evol}^{mono} (E_i) 
\end{align}

The relative probability of an idle state on a mono-filled site, $P_{idle}^{rel, mono} (E_i)$, can be evaluated as the mutually exclusive probability to $P_{evol}^{mono} (E_i)$, as given in Eq. \ref{eq: Pidle_on_mono}.

\begin{align} \label{eq: Pidle_on_mono}
    P_{idle}^{rel, mono} (E_i) & = 1 - \sum\limits_{event} P_{event}^{rel, mono}(E_i)  = 1 - P_{evol}^{mono} (E_i)  
\end{align}

Concerning the absolute probability of the evolution of the system for species $i$ on a site with BE $E_i$ concurrently filled by a species $j$ with BE $E_j$, it is defined in Eq. \ref{eq: Pevol_on_double}. As a working assumption, the BE of a species is assumed to be unaffected by the transient presence of a second adsorbate on the same BS. Accordingly, $E_i$ and $E_j$ are treated independently in double-filled configurations. The sole exception is the H$_2$ encounter-desorption mechanism, for which the BE is sharply reduced. This assumption is not fundamental and can be relaxed, as discussed in Appendix \ref{app: Encounter BE} and referenced where relevant throughout the following derivations.

\begin{align} \label{eq: Pevol_on_double}
    P_{evol}^{double} (E_i, E_j) & = 1 - \Big [(1 - P_{diff} (E_i))(1 - P_{des} (E_i)) \\ \notag
    & \times (1 - P_{diff} (E_j)) (1 - P_{des}(E_j) ) \\ \notag
    & \times (1 -  P_{cross, i+j}^{tot, excl}) \Big ] \\\notag
    P_{cross, i+j}^{tot, excl} &= 1 - \prod_{c_k}^{\substack{i+j \\ sub-channels}} (1 - P_{cross}^{i + j, c_k})
\end{align}

$P_{idle}^{rel, double} (E_i, E_j)$, the relative probability of an idle state of species $i$ on a double-filled BS, is defined as the mutually exclusive probability to $P_{evol}^{double} (E_i)$. The relative probabilities for diffusion, desorption, and reaction within double-filled configurations are given in Eqs. \ref{eq: Pdiff_eff_from_double}, \ref{eq: Pdes_eff_from_double}, and \ref{eq: Preac_eff_from_double}, respectively.

\begin{align}
    \label{eq: Pdiff_eff_from_double}
    &P_{diff}^{rel, double}  (E_i, E_j) = \frac{\nu_i(E_i) P_{diff} (E_i)}{W_{branching}^{double}(E_i,E_j)}\\ 
    \label{eq: Pdes_eff_from_double}
    &P_{des}^{rel, double} (E_i, E_j) = \frac{\nu_i(E_i) P_{des} (E_i)}{W_{branching}^{double}(E_i,E_j)}\\ 
    \label{eq: Preac_eff_from_double}
    &P_{reac, i+j}^{rel, double} (E_i, E_j) = \frac{\nu_{i,j}(E_i,E_j) P_{cross, i+j}^{tot, sum}}{W_{branching}^{double}(E_i,E_j)}
\end{align}

\noindent with

\begin{align}
    \notag
    W_{branching}^{double}(E_i,E_j) = \nu_{i,j}(E_i,&E_j) P_{cross, i+j}^{tot, sum} + \sum\limits_{k=i,j} \sum\limits_{event} \nu_k(E_k) P_{event}(E_k)
    \\
    \notag
    \sum\limits_{k=i,j} \sum\limits_{event} \nu_k(E_k) P_{event}(E_k) &= \nu_i(E_i) (P_{diff} (E_i) + P_{des} (E_i)) \\ \notag 
    &+ \nu_j(E_j) (P_{diff} (E_j) + P_{des} (E_j)) 
\end{align}

\noindent and

\begin{equation}
    \notag
    P_{cross, i+j}^{tot, sum} =  \sum_{c_k}^{\substack{i+j \\\notag sub-channels}} P_{cross}^{i+j, c_k}
\end{equation}

The relative probabilities from double filled-configurations are weighted by the characteristic initiation frequencies of the competing channels. For diffusion and desorption, the relevant clock is the trial frequency of the species that initiates the event, $\nu_i(E_i)$ or $\nu_j(E_j)$. For the reactive channel, an encounter-level reaction attempt frequency $\nu_{i,j}(E_i,E_j)$ is introduced, associated with the local oscillatory attempts of the two reactants to cross the reaction barrier. Following the same considerations as in \citet{2011ApJ...735...15G}, this frequency is taken as

\begin{equation} \label{eq: trial_frequency_reac}
\nu_{i,j}(E_i,E_j)=\max\left[\nu_i(E_i),\nu_j(E_j)\right], 
\end{equation}

\noindent so that the reaction clock is governed by the largest trial frequency of the encounter pair. The introduction of the trial frequencies into the relative probabilities restores the intrinsic initiation timescale associated with each species. Weighting the competing channels by their corresponding trial frequencies therefore ensures that both the probability of success of a process and its characteristic initiation frequency contribute to the competition. Consequently, species associated with larger trial frequencies are assigned a proportionally larger weight in the branching probabilities. 

The absence of explicit multiplication by the evolution probability in double-filled relative probabilities is worth highlighting. The probabilistic meaning nevertheless remains identical to that of mono-filled sites, as they still represent evolution-conditioned relative probabilities. This apparent difference arises from the subsequent treatment of the transient pseudo-idle state. Unlike the terminal idle state associated with mono-filled configurations, the pseudo-idle state belongs to the transient subspace and therefore does not correspond to a directly observable outcome at the ODE-solver level. Consequently, the pseudo-idle state contribution must be accounted for over its multiple possible revisits prior to any subsequent event. This is quantified through a geometric summation over $P_{idle}^{rel, double}$, given in Eq. \ref{eq:P_idle_double_geometric_sum},

    \begin{align}
    \label{eq:P_idle_double_geometric_sum}
    \sum_{n = 0} (P_{idle}^{rel, double} (E_i, E_j))^n &= \frac{1}{1 - P_{idle}^{rel, double} (E_i, E_j)} \\
     &= \frac{1}{P_{evol}^{rel, double} (E_i, E_j)} \notag
    \end{align}

\noindent effectively canceling out the explicit evolution factor appearing in the mono-filled formulation. To avoid unnecessarily burdening the following expressions, the multiplication by the evolution probability is therefore kept implicit. As discussed in the following subsections, this factor is naturally recovered during the treatment of repeated pseudo-idle revisits and the associated geometric-series formulation. The resulting treatment preserves the single-occupation constraint imposed at the ODE-solver level while consistently accounting for transient double-filled configurations and their intrinsic process competitiveness.

The relative probabilities introduced above provide a self-consistent framework to describe the competition between all accessible processes while preserving the correct branching of the associated probability fluxes. An immediate consequence of this competitive formalism concerns the H$_2$ encounter effect and its natural inclusion within the present formalism. When two H$_2$ occupy the same BS, the effective BE entering the diffusion and desorption probabilities is reduced to the encounter-desorption value (23~$K$ in \cite{Hincelin2015}). As a result, both the diffusion and desorption channels are simultaneously affected through their respective absolute probabilities, and their relative prevalence is self-consistently accounted for through the competitive branching formalism. As further formalized in the following, this coupling naturally captures the interplay between enhanced desorption and enhanced mobility. The resulting behavior is naturally accounted for from the competition between diffusion and desorption, without requiring any additional ad hoc treatment, in contrast to the pure RE-based method.
 
A binary switch, noted $s_{\rm barrierless}^{i+j}$, is introduced to distinguish between the competitive and non-competitive LH regimes (Eq. \ref{eq: switch_barrierless}). It takes a value of unity for barrierless channels and zero otherwise. In the barrierless limit, the reaction occurs immediately upon encounter, and the reactive channel directly reaches the quasi-absorbing reaction state. For activated reactions, by contrast, the reactive channel is a competitive branch within the TFC framework, requiring the explicit treatment of its competition with all other TFC-relevant processes and a proper branching of the resulting fluxes. The switch, therefore, naturally selects between direct reaction upon encounter and competitive barrier crossing while preserving a unified probabilistic formalism.

    \begin{equation}\label{eq: switch_barrierless}
    s_{\rm barrierless}^{i+j} = \begin{cases} 1, & P_{cross, i+j}^{tot, excl}=1,\\ 0, & \mathrm{otherwise}. \end{cases} \end{equation}

\subsubsection{Transient states subspace description}\label{sec: transient}

Upon entering the transient subspace, species may undergo successive hopping interspersed with pseudo-idle states before eventually reaching an absorbing state. In this scope,  $P_{\rm survival}^{\rm transient, eff,i}$ denotes the effective probability for a transient trajectory to remain within the transient subspace after an additional trial rather than reaching a (quasi-)absorbing state. As commented previously, the revisit multiplicity of pseudo-idle double-filled configurations is already embedded in the relative probabilities for double-filled configurations; $P_{\rm survival}^{\rm transient, eff, i}$ could consequently be viewed as an effective probability of transient subchain survival. $P_{\rm survival}^{\rm transient, eff, i}$, given in Eq. \ref{eq: P_eff_re-diff}, is constructed from the conditional probabilities of transient state survival from mono-filled and double-filled configurations, $P_{\rm re-diff}^{\rm eff, mono}$ and $P_{\rm re-diff}^{\rm eff, double}$, weighted by the corresponding probabilities $P^{mono}$ and $P^{double}$ of the next event initiation being from these respective configurations.

\begin{equation} \label{eq: P_eff_re-diff}
    P_{\rm survival}^{\rm transient, eff, i} = P^{\rm mono}P_{\rm survival, mono}^{\rm transient, eff, i} + P^{\rm double} P_{\rm survival, double}^{\rm transient,  eff, i}
\end{equation}

\noindent where $P^{\rm mono}P_{\rm survival, mono}^{\rm transient, eff, i}$ can be computed from

\begin{align}
    \notag
    & (1 - \Theta) \int \frac{(1 - \theta_i(E_i'))g_i(E_i')}{(1 - \Theta_i)} P_{diff}^{rel, mono} (E_i') \, dE_i'
\end{align}

\noindent while the definition of the double-filled contribution is given by 

\begin{align}
    \notag
     &P^{\rm double} P_{\rm survival, double}^{\rm transient,  eff, i}\\\notag
     &= \sum_{j \neq i} \int \theta_{j} (E_{j}) g_{j} (E_{j}) \int \frac{(1 - \theta_i (E_i'))g_i(E_i')}{(1 - \Theta_i)} \\\notag
     & \times (1 - s_{\rm barrierless}^{i+j}) \biggl[P_{diff}^{rel, double} (E_i', E_j)  \\\notag
     &+ (P_{diff}^{rel, double}(E_j, E_i') + P_{des}^{rel, double} (E_j, E_i')) P_{diff}^{rel, mono}(E_i')\biggl] dE_i \, dE_j \\ \notag
     & + \int \theta_i (E_i') g_i (E_i') \times (1 - s_{\rm barrierless}^{i+i}) \times \biggl[P_{diff}^{rel, double} (E_i', E_i') 
     \\\notag &+ (P_{diff}^{rel, double} (E_i', E_i') + P_{des}^{rel, double} (E_i', E_i')) P_{diff}^{rel, mono}(E_i')\biggl] \, dE_i
\end{align}

\noindent where the first double integral accounts for double-filled configurations for $i+j$ pairs with $j$ different from $i$. The second simple integral accounts for the cases where the filling species is a species $i$. We note that if the diffusion/desorption of $j$ occurs first, then the species $i$ is left alone at its BS; the relative probability of subsequent diffusion must be evaluated using the mono-filled definition, as given in the last term of both integrals.

The microscopic sequence of transient states is not explicitly resolved. Instead, its cumulative effect is captured statistically through the effective transient state survival probability. More specifically, analogously to the geometric summation over the pseudo-idle state discussed above (Eq. \ref{eq:P_idle_double_geometric_sum}), whose effect is implicitly incorporated into the normalization of the double-filled relative probabilities entering the definition of $P_{\rm survival}^{\rm transient,  eff, i}$, repeated returns to the transient subspace are accounted for through a geometric summation over $P_{\rm survival}^{\rm transient, eff, i}$. This summation quantifies the expected multiplicity of transient revisits prior to irreversible absorption, as expressed in Eq. \ref{eq:P_rediff_geometric_sum}.  

    \begin{equation}
    \label{eq:P_rediff_geometric_sum}
    \sum_{n = 0} (P_{\rm survival}^{\rm transient,  eff, i})^n = \frac{1}{1 - P_{\rm survival}^{\rm transient, eff, i}} 
    \end{equation}

This multiplicity factor does not specify how the transient revisits are distributed among the accessible transient configurations. The corresponding statistical weighting is determined from the BED-integrated transient entry probability together with the species-resolved coverages and vacancy fractions associated with the current state. Their combination defines the quasi-stationary distribution (QSD) of the transient subspace, which describes the statistical distribution of transient configurations visited prior to absorption. The QSD provides the complete statistical description required to evaluate the cumulative contribution of transient trajectories originating from a given current state without resolving their microscopic sequence. This, therefore, provides the deterministic closure of the transient dynamics. The resulting formalism is presented in the following subsection.

\subsubsection{TFC ODE-level description: BE-resolved stationary distribution of Markov states and net event rate}

\begin{figure*}
    \centering
    \includegraphics[width=0.9  \linewidth]{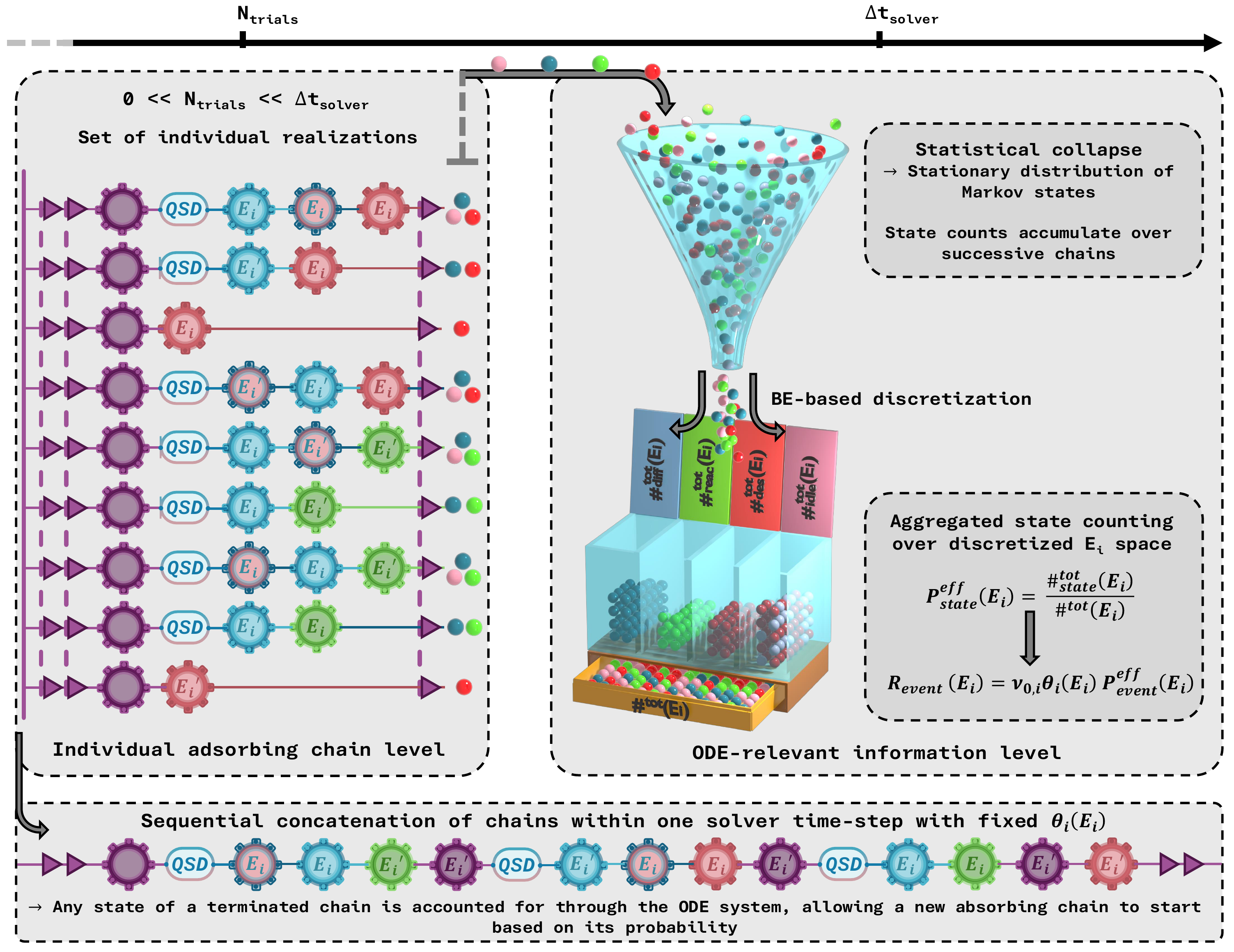}
    \caption{Schematization of the concatenation procedure towards the stationary distribution of Markov state defining the ODE-relevant information level. The central schematic elements (collapse funnel and state box) were designed using Blender version 4.0.1. }
    \label{fig:counting_explanation}
\end{figure*}

Fig. \ref{fig:counting_explanation} illustrates the conversion from the absorbing Markov-chain formalism into the BE-resolved rates required for the ODE system definition of the $\theta_i(E_i)$ time-derivative. During a single solver time step, the BE-resolved coverages $\theta_i(E_i)$ are constant. Since the characteristic number of trials required for chain termination is much smaller than the number of trials occurring within one solver time step, numerous absorbing-chain realizations can be successively generated under fixed BE-resolved species coverages. The concatenation of these realizations corresponds to a statistical collapse of the ensemble of possible chains, weighted by their respective probabilities of occurrence. The microscopic ordering of individual realizations is not retained; only the statistical weight associated with each Markov state on BS $E_i$ is preserved. An equivalent mathematical formulation is obtained by reconnecting the absorbing states of the Markov chain to the current state. The resulting ergodic chain possesses a non-trivial stationary distribution over all Markov states, including the transient configurations. This stationary distribution is mathematically equivalent to the absorbing chain concatenation procedure and provides the statistical weights associated with the Markov state BE-resolved stationary occupancies. Fig. \ref{fig:counting_explanation} illustrates these two complementary views of the same construction.  

The mathematical implementation of the concatenation procedure requires the evaluation of the occurrence probabilities of each BE-resolved Markov state associated with the different subbranches of absorbing-chain realizations. In the particular case of subchains passing through the transient subspace, the corresponding statistical weights are described by the QSD introduced in Sect. \ref{sec: transient}. The construction of this QSD requires the BED-integrated effective probability for a species $i$ to enter the transient subspace from the current state through the mandatory diffusive gateway, noted $P_{\rm diff, \rm gateway}^{eff, i}$, as given in Eq. \ref{eq: coutingevent_start_diff}. 

\begin{equation} \label{eq: coutingevent_start_diff}
    P_{\rm diff, \rm gateway}^{eff, i} = \int \theta_i (E_i') g_i (E_i') P_{diff}^{rel, mono} (E_i') \, dE_i'
\end{equation}

The BE-resolved occurrence counts combine direct contributions from the current state and contributions originating from transient-mediated subchains. The total diffusion count on BSs of energy $E_i$, $\#_{diff}^{tot}(E_i)$ is given by Eq. \ref{eq: coutingevent_tot_diff_Ei}.

\begin{align}\label{eq: coutingevent_tot_diff_Ei}
    \#_{diff}^{tot} (E_i) &= \#_{diff}^{start} (E_i) + \#_{diff}^{transient} (E_i) 
\end{align}

\noindent where $\#_{diff}^{start} (E_i)$ and $\#_{diff}^{transient} (E_i)$ are, respectively, the counts of starting diffusion from the current state and transient diffusion states on BS $E_i$, as

\begin{align}
    \#_{diff}^{start} (E_i) &= \frac{1}{\nu_{i, 0}(E_i)} \theta_i (E_i) P_{diff}^{rel, mono} (E_i) \notag
\end{align}

\noindent and

\begin{align}
    \notag
    \#&_{diff}^{transient} (E_i) = \frac{1}{\nu_{i, 0}(E_i)} \times \frac{P_{\rm diff, \rm gateway}^{eff, i}}{(1 - P_{\rm survival}^{\rm transient, eff, i})} \Bigg[ \\ \notag
     & (1 - \Theta) \frac{(1 - \theta_i (E_i))}{(1 - \Theta_i)} P_{diff}^{rel, mono} (E_i)\\ \notag
     & + \sum_{j \neq i} \Bigg( \int \theta_j (E_j)g_j (E_j) \frac{(1 - \theta_i (E_i))}{(1 - \Theta_i)} (1 - s_{\rm barrierless}^{i+j}) \times \\ \notag
     &  \biggl[P_{diff}^{rel, double} (E_i, E_j) \\\notag
     &+ (P_{diff}^{rel, double}(E_j, E_i) + P_{des}^{rel, double} (E_j, E_i)) P_{diff}^{rel, mono}(E_i)\biggl] dE_j \Bigg) \\ \notag
     & + \theta_i (E_i) (1 - s_{\rm barrierless}^{i+i}) \times \biggl[P_{diff}^{rel, double} (E_i, E_i) 
     \\\notag 
     &+ (P_{diff}^{rel, double} (E_i, E_i) + P_{des}^{rel, double} (E_i, E_i)) P_{diff}^{rel, mono}(E_i)\biggl] \Bigg]
\end{align}

We note that in the case of $\#_{diff}^{transient} (E_i)$, the occupancy/vacation-related terms multiplied by the multiplicity of transient revisit, weighted by the effective probability of entrance to the transient state, account for the transient-mixing QSD. This comment holds for each of the following counts for internal or terminal states of a transient subchain.   

In the case of idle states, we defined three forms; these are

\begin{itemize}
    \item[(a)] the starting event of the Markov chain from the current state with direct access to the absorbing idle state 
    \item[(b)] an internal transient state after diffusion on an occupied site, with a sequence of transient, pseudo-idle states on double-filled BSs, before any subsequent event (diffusion or subchain termination)
    \item[(c)] a terminal state of the transient subspace, restricted to effective mono-filled BSs.
\end{itemize}

\noindent The total count for (pseudo-)idle states for species $i$ on a site $E_i$ is consequently given in Eq. \ref{eq: coutingevent_tot_idle_Ei}. 

\begin{equation}
    \label{eq: coutingevent_tot_idle_Ei}
    \#_{idle}^{tot} (E_i) = \#_{idle}^{start} (E_i) + \#_{idle}^{transient/terminal} (E_i)
\end{equation}

\noindent with $\#_{idle}^{start} (E_i)$ describing the counting of idle states of type (a)

\begin{equation}
    \#_{idle}^{start} (E_i) = \frac{1}{\nu_{i, 0}(E_i)} \theta_i (E_i) P_{idle}^{rel, mono} (E_i)\notag
\end{equation}

\noindent while $\#_{idle}^{transient/terminal} (E_i)$ quantifies the counting of transient pseudo-idle states (type b) and terminal idle states ending transient subchains (type c defined above)

\begin{align}
    \notag
    \#&_{idle}^{transient/terminal} (E_i) = \frac{1}{\nu_{i, 0}(E_i)} \times \frac{P_{\rm diff, \rm gateway}^{eff, i}}{(1 - P_{\rm survival}^{\rm transient,  eff, i})} \Bigg[ \\ \notag
     & (1 - \Theta) \frac{(1 - \theta_i (E_i))}{(1 - \Theta_i)} P_{idle}^{rel, mono} (E_i)\\ \notag
     & + \sum_{j \neq i} \Bigg( \int \theta_j (E_j)g_j (E_j) \frac{(1 - \theta_i (E_i))}{(1 - \Theta_i)} (1 - s_{\rm barrierless}^{i+j}) \times \\ \notag
     & \Bigg( \frac{P_{idle}^{rel,double} (E_i, E_j)}{P_{evol}^{double} (E_i, E_j)} \\ \notag
     & + (P_{diff}^{rel, double}(E_j, E_i) + P_{des}^{rel, double} (E_j, E_i)) P_{idle}^{rel, mono}(E_i) \, dE_j \Bigg) \\ \notag
     & + \theta_i (E_i) (1 - s_{\rm barrierless}^{i+i}) \times \Bigg( \frac{P_{idle}^{rel,double} (E_i, E_i)}{P_{evol}^{double} (E_i, E_i)} \\ \notag
     & + (P_{diff}^{rel, double} (E_i, E_i) + P_{des}^{rel, double} (E_i, E_i)) P_{idle}^{rel, mono} (E_i) \Bigg) \Bigg]
\end{align} 

\noindent where the contribution (b) associated with a pseudo-idle configuration $P_{idle}^{rel, double}(E_i,E_j)$ is weighted by its revisit multiplicity prior to a subsequent evolutionary event, consistent with the normalization previously introduced for the relative probabilities of double-filled configurations.

In the case of desorption events, The total counts from species $i$ on site with BE $E_i$ are defined by Eq. \ref{eq: coutingevent_tot_des_Ei}. 

\begin{align}
    \label{eq: coutingevent_tot_des_Ei}
    \#_{des}^{tot} (E_i) = \#_{des}^{start} (E_i) + \#_{des}^{terminal} (E_i)
\end{align}

\noindent with the direct desorption contribution from the current state given by

\begin{align}
    \#_{des}^{start} (E_i) = \frac{1}{\nu_{i, 0}(E_i)} \theta_i (E_i) P_{des}^{rel, mono} (E_i) \notag
\end{align}

\noindent while the terminal contribution after passing through the transient subspace is expressed as

\begin{align}
    \notag
    \#&_{des}^{terminal} (E_i) = \frac{1}{\nu_{i, 0}(E_i)} \times \frac{P_{\rm diff, \rm gateway}^{eff, i}}{(1 - P_{\rm survival}^{\rm transient, eff, i})} \Bigg[ \\ \notag
     & (1 - \Theta) \frac{(1 - \theta_i (E_i))}{(1 - \Theta_i)} P_{des}^{rel, mono} (E_i)\\ \notag
     & + \sum_{j \neq i} \Bigg( \int \theta_j (E_j)g_j (E_j) \frac{(1 - \theta_i (E_i))}{(1 - \Theta_i)} (1 - s_{\rm barrierless}^{i+j}) \times \\ \notag
     &  \biggl[P_{des}^{rel, double} (E_i, E_j)\\\notag
     &+ (P_{diff}^{rel, double}(E_j, E_i) + P_{des}^{rel, double} (E_j, E_i)) P_{des}^{rel, mono}(E_i)\biggl] dE_j \Bigg) \\ \notag
     & + \theta_i (E_i) (1 - s_{\rm barrierless}^{i+i}) \times \biggl[P_{des}^{rel, double} (E_i, E_i) 
     \\\notag 
     &+ (P_{diff}^{rel, double} (E_i, E_i) + P_{des}^{rel, double} (E_i, E_i)) P_{des}^{rel, mono}(E_i)\biggl]
\end{align}

Finally, the counting of reaction events is given in Eq. \ref{eq: coutingevent_tot_reac_Ei}. Actually, the TFC-capped reactive channels only concern the LH-like mechanism, as previously commented, and therefore define the reactive terminal path transient subchain. 

\begin{equation}
    \label{eq: coutingevent_tot_reac_Ei}
    \#_{reac, i+j}^{tot} (E_i) = \#_{reac, i+j}^{terminal} (E_i)
\end{equation}

\noindent where $\#_{reac,i+j}^{terminal}(E_i)$, the counting of the total terminal reaction of $i$ and $j$ on double-filled sites is given if $j \neq i$ by

\begin{align}
    \notag
    \#_{reac, i+j}^{terminal} (E_i) &=  \frac{P_{\rm diff, \rm gateway}^{eff, i}}{(1 - P_{\rm survival}^{\rm transient, eff, i})} \\ \notag
     & \int \frac{1}{\nu_{i,j}(E_i,E_j)} \theta_j (E_j)g_j (E_j) \frac{(1 - \theta_i (E_i))}{(1 - \Theta_i)} \\ \notag
     & \{ s_{\rm barrierless}^{i+j} + (1 - s_{\rm barrierless}^{i+j}) P_{reac, i+j}^{rel, double} (E_i, E_j) \} \, dE_j 
\end{align}

\noindent if $j = i$, then

\begin{align}
    \notag
    \#_{reac, i+i}^{terminal} (E_i) &= \frac{1}{\nu_{i,i}(E_i,E_i)} \times \frac{P_{\rm diff, \rm gateway}^{eff, i}}{(1 - P_{\rm survival}^{\rm transient,  eff, i})} \\\notag
     & \theta_i (E_i) \{ s_{\rm barrierless}^{i+i} + (1 - s_{\rm barrierless}^{i+i}) P_{reac, i+i}^{rel, double} (E_i, E_i)\} 
\end{align}

The dynamical counterpart for reaction events from $j$ species landing on a site filled by species $i$ is defined in Eq. \ref{eq: coutingevent_tot_reac_Ej_->_E_i}. 

\begin{equation}
    \label{eq: coutingevent_tot_reac_Ej_->_E_i}
    \#_{reac, j+i}^{tot} (E_j \rightarrow E_i) = \#_{reac, j+i}^{terminal} (E_j \rightarrow E_i)
\end{equation}

In the case where $j \neq i$, $\#_{reac, j+i}^{terminal} (E_j \rightarrow E_i)$ is expressed as

\begin{align}
    \notag
    \#&_{reac, j+i}^{terminal} (E_j \rightarrow E_i)  = \frac{1}{\nu_{i,j}(E_i,E_j)} \times \frac{P_{\rm diff, \rm gateway}^{eff, j}}{(1 - P_{\rm survival}^{\rm transient, eff, j})}  \\ \notag
     & \theta_i (E_i) \frac{(1 - \theta_j (E_j))}{(1 - \Theta_j)} \Bigg(s_{\rm barrierless}^{i+j} + (1 - s_{\rm barrierless}^{i+j}) P_{reac, j+i}^{rel, double} (E_i, E_j) \Bigg)
\end{align}

while, if $j = i$, it reduces to 

\begin{align}
    \notag
    \#&_{reac, i+i}^{terminal} (E_i \rightarrow E_i) = \#_{reac, i+i}^{terminal} (E_i)
\end{align}

The counting of the total number of Markov states implying species $i$ on a site with BE $E_i$ is then given in Eq. \ref{eq: coutingevent_tot_on_Ei}

\begin{align}
    \label{eq: coutingevent_tot_on_Ei}
    \#^{tot} (E_i) = \sum_{state} \#_{state}^{tot} (E_i) &= \#_{diff}^{tot} (E_i) + \#_{idle}^{tot} (E_i) + \#_{des}^{tot} (E_i) \\\notag & + \sum_{j \neq i}\#_{reac, i+j}^{terminal} (E_i) + \#_{reac, i+i}^{terminal} (E_i)
\end{align}

The effective probability of each Markov state for species $i$ on a BS with BE $E_i$ is defined as its statistical weight within the BE-resolved stationary occupancies. It is expressed by Eq. \ref{eq:P_eff_event_E_i}.

\begin{equation}
    \label{eq:P_eff_event_E_i}
    P_{state}^{eff} (E_i) = \frac{\#_{state}^{tot} (E_i)}{\#^{tot} (E_i)}
\end{equation}

We note that the dynamical counterpart of reactive channels for which a species $j$ is landing on a species $i$ on a BS with BE $E_j$, $E_i$ is limited by the trial frequency and dynamical budget of species $j$ on BS $E_j$. The effective probability of a $j+i$ reaction from species $j$ landing on a species $i$ on a BS with BE $E_j$, $E_i$ is therefore given by Eq. \ref{eq:P_eff_reac_E_j_->_E_i}. 

\begin{equation}
    \label{eq:P_eff_reac_E_j_->_E_i}
    P_{reac, j+i}^{eff} (E_j \rightarrow E_i) = \frac{\#_{react, j+i}^{tot} (E_j \rightarrow E_i)}{\#^{tot} (E_j)}
\end{equation}

The corresponding BE-resolved net event rates are then defined in Eqs. \ref{eq: RateEvent_i_E_i} and \ref{eq: RateReact_E_j_->_E_i}. Only coverage-affecting states (diffusion, desorption, and reaction) are considered, as idle states leave the surface coverages unchanged and therefore carry no contribution to the ODE-level temporal evolution. These rates are constructed by multiplying the effective event probability with the corresponding trial frequencies and BE-specific coverages, thereby recovering the canonical structure of rate expressions. 

\begin{align} \label{eq: RateEvent_i_E_i}
    R_{event} (E_i) &= \nu_{i,0}(E_i) \theta_i (E_i) P_{state}^{eff} (E_i)
\end{align}

The trial frequency appearing in Eq. \ref{eq: RateEvent_i_E_i} is fundamentally dictated by the characteristic frequency at which a microscopic event occurs within a BS of BE $E_i$. However, since local chemical reactions (Eq. \ref{eq: trial_frequency_reac}) and encounter mechanisms can dynamically alter the localized trial frequency of the system, this parameter must be transformed into an event-dependent effective trial frequency, as derived explicitly in Appendix \ref{app: Encounter BE}. The reactive counterpart from $j$ diffusing on $i$ is given by Eq.~\ref{eq: RateReact_E_j_->_E_i}.

\begin{align} \label{eq: RateReact_E_j_->_E_i}
    R_{reac, j+i} (E_j\rightarrow E_i) &= \nu_{i,j}(E_i,E_j) \theta_j (E_j) P_{reac, j+i}^{eff} (E_j\rightarrow E_i)
\end{align}

The last equation can finally be integrated over $E_j$, as expressed in Eq. \ref{eq: RateReact_E_i_jpi}, in order to prepare its subsequent inclusion within the contributions to the $\theta_i(E_i)$ time derivatives. 

\begin{equation}\label{eq: RateReact_E_i_jpi}
    R_{reac, j + i} (E_i) = \int R_{reac, j + i} (E_j, E_i) g_j (E_j) \, dE_j
\end{equation}

Branching ratios for the detailed contributions of the microscopic pathways underlying a given Markov state are provided in Appendix \ref{app: BR}. These include the contributions associated with the different thermal equilibrium states, as well as the independent tunneling-mediated channels when applicable. Such branching ratios are not required for the ODE integration itself, which only depends on the total contribution associated with a given coverage-affecting Markov state. They are nevertheless useful for output pre-processing purposes, allowing the individual microscopic pathways contributing to a given event to be reconstructed, visualized, and quantitatively analyzed.

\subsection{Coupled ODE system formulation}
\label{Sect: ODE system formulation}

In the previous subsection, we introduced an absorbing Markov chain-based treatment of the TFC system. However, processes that are not limited by a trial frequency must be incorporated separately. These include ER reactions, accretion from the gas phase, and ice processing driven by non-thermal mechanisms such as CR-sputtering and (secondary) UV photon-induced desorption.

The ER reaction rates between a solid-phase species $i$ and a gas-phase species $j$, $R_{i + j}^{ER} (E_i)$, can be computed straightforwardly using Eq. \ref{eq: R_ER_i+j}. 

\begin{equation}\label{eq: R_ER_i+j}
    R_{reac, j + i}^{ER} (E_i) = \theta_i (E_i) S_j \frac{\sigma_{gr}}{N_{site}} v_{th, j} n_j P_{cross, j+i}^{tot, excl}
\end{equation}

\noindent where $\sigma_{gr}$ is the grain cross-section ($= \pi r_{grain}^2$), $S_j$ is the sticking coefficient $[0-1]$, $v_{th, j}$ is the thermal velocity, and $n_j$ is the number density of the gas phase species $j$. We note that, in cases where the gas and grain species are not characterized by the same temperature ($T_{gas} \ne T_{dust}$), the temperature used in $P_{cross}^{i + j}$ is computed via Eq. \ref{eq: T_eff_ER} \citep{2013A&A.559.A49.9},

\begin{equation}\label{eq: T_eff_ER}
    T_{eff}^{ER} = \mu \left(\frac{T_{dust}}{m_{i_s}}+\frac{T_{gas}}{m_{j_g}}\right) 
\end{equation}

Similarly, the accretion rate on a BS with BE $E_i$ for species $i$, $R_{acc}(E_i)$ can be computed through Eq. \ref{eq: R_accretion}. 

\begin{equation}\label{eq: R_accretion}
    R_{acc}(E_i) = (1-\Theta) \frac{(1 - \theta_i (E_i))}{(1 - \Theta_i)} S_i \frac{\sigma_{gr}}{N_{site}} v_{th, i} n_i
\end{equation}

The net gain rate for $\theta (E_i)$ can then be computed via Eq. \ref{eq: R_gain}. 

\begin{align}\label{eq: R_gain}
    R^{gain} &(E_i) = R_{acc}(E_i) + \frac{(1 - \theta_i (E_i))}{(1 - \Theta_i)} \times \int R_{diff} (E_i') g_i (E_i') \, dE_i' \notag\\
    & + 
    \sum_{\substack{j + k \rightarrow i + ... \\ j \neq k}}
    \bigg[
    \big(1 - f_{cd}^{j + k} (E_i)\big)
    \times \alpha_{j + k \rightarrow i}^{stoechio}
    \frac{P_{cross}^{j + k \rightarrow i + ...}}{P_{cross, j+k}^{tot, sum}}
    \\
     & \times
     \bigg(
     \int (R_{reac, j + k} (E_j) + R_{reac, j + k}^{ER} (E_j))
     \times g_j (E_j) \, dE_j
     \notag\\
     & +
     \int (R_{reac, k + j} (E_k) + R_{reac, k + j}^{ER} (E_k))
     \times g_k (E_k) \, dE_k
     \bigg)
     \bigg]
     \notag\\
     & +
     \frac{(1 - \theta_i (E_i))}{(1 - \Theta_i)} \times \sum_{j + j \rightarrow i + ...}
     \bigg[
     \big(1 - f_{cd}^{j + j} (E_i)\big)
     \times \alpha_{j + j \rightarrow i}^{stoechio}
     \frac{P_{cross}^{j + j \rightarrow i + ...}}{P_{cross, j+j}^{tot, sum}}
     \notag\\\notag
     & \times     
     \int (R_{reac, j + j} (E_j) + R_{reac, j + j}^{ER} (E_j))
     \times g_j (E_j) \, dE_j
     \bigg]
\end{align}

\noindent where $\alpha_{j + k \rightarrow i}^{stoechio}$ is the stoichiometric coefficient of the product $i$. The ratio $\frac{
P_{cross}^{i+j,c_k}
}{
P_{cross,i+j}^{tot,sum}
}$ accounts for the branching ratio of reactions originating from the same reactive pair, leading to a distinct set of products, as further discussed in Appendix \ref{app: BR}. $f_{cd}^{j + k} (E_i)$ quantifies the chemical desorption efficiency, ranging between 0 and 1; it therefore defines an absolute probability of desorption induced by an exothermic surface reaction (Sect. \ref{Sect: Chemical Desorption Efficiency}). We note that chemical desorption is only considered for reaction products that have just formed. We therefore neglect the possibility that a juxtapozed species benefits from the excess energy of the reaction. We also note that chemical diffusion is not considered, since it would involve an iterative problem-solving process with constraints on the dissipation time before thermalization with the surface.

It is worth highlighting the $j+j$ contribution, treated separately due to the indistinguishability of identical-reactant encounters and the associated mass-conservation constraint. This is equivalent to the usual $(1/2)$ symmetry correction employed in classical rate equations, which is implicitly compensated here by a factor $(2)$ associated with the two identical $R_{reac, j + j}^{(ER)} (E_j)$ and related dynamical counterpart gain contributions. 

Concerning the loss term, the remaining contribution to be defined is the non-thermal desorption (Eq. \ref{eq: R_non-thermal_des}). In the present formulation of a monolayer, energetic processing is restricted to enhanced desorption, including CR-sputtering and photodesorption. The internal contributions are detailed in Appendix \ref{app: ice_proc}. 

\begin{equation}\label{eq: R_non-thermal_des}
    R_{des}^{non-th, tot}(E_i) = \sum_{X=[\rm CR-sput, UV]} R_{des}^{non-th, X}(E_i)
\end{equation}

$R^{loss}(E_i)$ is then obtained through Eq. \ref{eq: R_loss}.

\begin{align}\label{eq: R_loss}
    R^{loss} (E_i) &=  R_{diff} (E_i) + R_{des} (E_i) + R_{des}^{non-th, tot}(E_i) \\\notag
    &+ \sum_{j} \biggl[ R_{reac, i + j} (E_i) + R_{reac, j + i}^{ER} (E_i) + R_{reac, j + i} (E_i) \biggl]
\end{align}

Finally, the final coupled ODE system is simply given by 

\begin{equation}\label{eq: ODE}
    \frac{d \theta (E_i)}{dt} = R^{gain} (E_i) - R^{loss} (E_i)
\end{equation}

The associated gas-phase gain and loss coupling terms are derived from the integration of Eqs. \ref{eq: R_gain} and \ref{eq: R_loss} over the species $i$ BED, scaled by both $N_{site}$ and the dust grain number density to consistently recover coupling terms expressed in gas-phase number densities. Their expressions are given in Appendix \ref{app: gas_coupling}.

\section{DESTINY numerical code}
\label{Sect: Numerical_code}

Since astrochemical networks inherently constitute highly stiff ODE systems, the implementation of Backward Differentiation Formulae methods is requisite to numerically simulate the temporal evolution of molecular abundances. To this end, the CVODES solver from the SUNDIALS suite \citep{hindmarsh2005sundials, gardner2022sundials, cvodesDocumentation} is coupled with the newly developed DESTINY architecture in Modern Fortran, which computes the right-hand-side vector of the governing ODE problem (Eq. \ref{eq: ODE}). In contrast to the standard CVODE package, the CVODES variant natively integrates sensitivity analysis capabilities, thereby providing pertinent diagnostics regarding parametric dependencies and model constraints.

As detailed in Sect. \ref{Sect: BED representation} – \ref{Sect: Chemical Desorption Efficiency}, DESTINY numerically resolves the mathematical framework delineated in the preceding section. Furthermore, the software architecture incorporates modular switches designed to activate or deactivate diverse numerical resolution techniques and astrochemical processes; a comprehensive description of these operational switches and their associated inputs is provided in Appendix\,\ref{app: Switches}.

\subsection{Binding energy distribution representation} \label{Sect: BED representation}

An advantage of the DESTINY framework is the absence of a priori assumptions regarding the form of the BED. In the code architecture, this flexibility is achieved by parameterizing any input BED as a linear combination of Gaussian functions. The resulting continuous BED is subsequently discretized onto bins via one of the two alternative discretization schemes, as chosen by the user: (i) maintaining a uniform resolution $\Delta E$ across all simulated chemical species, wherein the BED of a given species $i$ is partitioned into $n_i$ discrete bins, or (ii) enforcing a constant number of $n$ bins for all species, which conversely necessitates an adaptive resolution $\Delta E_i$. Since true Gaussian distributions possess infinite support, they must be truncated at their boundaries and renormalized prior to numerical resolution. Consequently, boundaries are established at a distance of $n_{sigma} \times \sigma$ from the mean, where $n_{sigma}$ is a user-defined parameter. For the case of a Gaussian mixture, the definitive global limits of the BED are computed via Eq. \ref{eq: truncation}. While the majority of current BEDs are well described by single Gaussian profiles, recent investigations show that some BEDs are better captured by double-Gaussian combinations, as exemplified by NH$_3$ \citep{Tinacci, 2025A&A...698A.284G}.

\begin{align} \label{eq: truncation}
    E_{min} &= min( \mu_k - n \times \sigma_k) ;\,\,\
    E_{max} = max( \mu_k + n \times \sigma_k)
\end{align}

\noindent for a mixture of $k$ Gaussians, where $\mu_k$ and $\sigma_k$ are respectively the mean and the standard deviation of the $k^{th}$ Gaussian. The absolute weight of each individual bin is subsequently evaluated via the complementary error function, a choice specifically implemented to mitigate numerical instabilities arising from bins characterized by vanishingly small statistical weights. To ensure conservation, these values are divided by the sum of all bin weights across the entire distribution. Following this normalization step, the effective energy of a given bin—required for the explicit BE-discretization of the trial frequency of species $i$—is obtained by integrating the energy-weighted BED over the bin domain and normalizing by the corresponding bin weight.

\subsection{Integration methods}

Numerical integration methods are of central importance within the DESTINY architecture, owing to their frequent recurrence arising from the BED discretization framework. Since TFC-event absolute probabilities are described through exponentially decreasing functions that constitute strictly convex functions, Jensen’s inequality \citep{Jensen1906} demonstrates that evaluating them using the effective energy value (i.e., the energy weighted by the BED) of each individual bin inherently underestimates the true integral values. To circumvent this underestimation, a change of variables is systematically implemented whenever feasible to cast the equations into the standard form $A \times \int_a^b e^{-t^2} \, dt$, thereby permitting an exact analytical solution via either the standard or complementary error function, the latter being favored in order to mitigate numerical noise for low-weight bins, as previously stated. Conversely, in instances where the governing equations cannot be mapped onto this analytical form—such as the bin-integrated absolute diffusion probability $P_{\text{diff}}^{T_x} (E_{bin_i})$, the absolute quantum tunneling diffusion probability evaluated over the specific energy domain $[E_i' \mid E_i' < E_i]$, as well as the subsequent overall integration of the absolute quantum tunneling diffusion probability over the source bin $P_{\text{diff}}^{\text{tunn}} (E_{bin_i})$— an adaptive Gauss-Legendre quadrature scheme is deployed to numerically approximate the integral, as described in Appendix \ref{app: Gauss_Legendre}.

\subsection{State vector projection under physical constraints}

During the numerical integration, small deviations from the physically admissible coverage domain may arise, particularly when the system evolves close to the coverage bounds or the monolayer saturation limit. To guarantee physically meaningful BE-resolved coverages throughout the integration, a dedicated projection function was implemented and supplied to the CVODES projection interface, allowing the state vector to be projected onto the admissible coverage domain after each accepted time step. This custom projection enforces both the individual coverage bounds and the global saturation constraint while preserving the closest admissible state under the Euclidean norm. It also improves the robustness and convergence of the integration by preventing unphysical coverages from entering and propagating through the RHS of the ODE system. The resulting constrained projection problem is solved through a combination of analytical and numerical procedures, ensuring machine-precision satisfaction of all physical constraints while minimizing computational overhead. A detailed mathematical derivation of the projection formalism is provided in Appendix \ref{appendix: projection_formalism}.

\subsection{The chemical desorption efficiency}
\label{Sect: Chemical Desorption Efficiency}

Analogous to the treatment of the trial frequency, the chemical desorption efficiency, $f_{\text{cd}}^{j + k}$, is parameterized via three alternative configurations: (i) a unique factor specified for each individual reaction within the grain-surface network; (ii) multiple user-defined coefficients explicitly mapped to each reaction pathway; or (iii) the predictive formalism prescribed by \cite{2016A&A...585A..24M}, which utilizes their parameterized formulation for kinetic energy transfer efficiency (Eq. \ref{eq: f_cd}).

\begin{equation} \label{eq: f_cd}
    f_{\text{cd}}^{j + k} (E_i) = e^{- \frac{E_i}{\epsilon E_{\rm Reac}^{\rm excess} / N}}
\end{equation}

\noindent  where $E_{\mathrm{Reac}}^{\mathrm{excess}} = min(0, -\Delta H_R)$ denotes the excess reaction energy. It takes non-zero values for exothermic channels. It is obtained from the reaction enthalpy and corresponds to the difference between the total formation enthalpies of the reactants and products. The parameter $N$ denotes the total number of degrees of freedom, defined by the relation $N = 3 \times n_{\text{atoms}}$, while $\epsilon$ is defined by the fraction of the liberated reaction energy retained by the desorbing adsorbate. This energy retention factor equals $\epsilon = \frac{(M - m)^2}{(M + m)^2}$, with $m$ the adsorbate mass and $M$ the effective surface mass, as prescribed in \cite{2016A&A...585A..24M}. Here, $M$ is currently set to $120$ amu according to the silicate-based value inferred by \cite{2016A&A...585A..24M}. Some specific reactions necessitate distinct chemical desorption efficiencies. These reaction-dependent factors are integrated directly within the DESTINY architecture, as outlined in Appendix \ref{app: ice_proc}. Furthermore, since the chemical desorption formalism introduced by \cite{2016A&A...585A..24M} is conceptually tailored to monoproduct reaction channels, DESTINY accommodates multiproduct pathways by permitting the optional integration of independent, user-defined efficiency factors for these reactions.

\section{Results and discussions}
\label{sec: result}

Results presented throughout this section are obtained using a reduced surface chemical network comprising 27 solid-phase species and 45 surface reactions. The complete list of solid-phase species and reactions included in the benchmark network is provided in Appendix \ref{app: benchmark_network}. This reduced network was constrained to the key solid-phase species in order to keep the interpretation of the results tractable and ease the assessment of the behavior of the proposed framework. We note that the gas-phase network comprises 584 species for 7667 reactions, as taken from \cite{2024A&A...689A..63W}. Simulations are performed for a cloud with $n_{\rm H}=3\times10^{4}$ cm$^{-3}$, $T_{gas}= T_{dust} = 12$ K, $A_V=15$, and $\zeta_{\rm CR}=1.3\times10^{-17}$ s$^{-1}$, adopting the same initial gas-phase abundances as in \citet{2016MNRAS.459.3756R}. Reaction and diffusion barrier widths for tunneling-assisted barrier-crossing are respectively set to 1 and 2.5 \AA. A $\chi$ of $0.4$ is adopted for all species. The input parameter list used in this paper is given in Appendix \ref{app: Switches}.

\subsection{Benchmarking DESTINY within the single-BE limit: highlights on the \texorpdfstring{H$_2$}{H2} encounter effect}\label{subsec: result_comp_to_Naut}

Before investigating the BED-induced effects, it is first necessary to assess the impact of the TFC/Markov-chain formalism itself and the resulting effective event probabilities derived from the stationary distribution of the Markov states. A benchmark against the open-source code Nautilus \citep{2016MNRAS.459.3756R, 2021A&A...652A..63W} is thereby carried out in the single-BE limit, where each species is represented by a single BE bin. This removes any effects associated with BED discretizations, allowing the consequences of the probabilistic reformulation alone to be evaluated. For this comparison, both models employ the same reduced surface network presented in Appendix \ref{app: benchmark_network}, and the same gas-phase network, disabling ER reactions. For atomic H, $\chi = 0.34$ is used in order to reproduce the diffusion barrier employed in Nautilus and ensure consistency with the benchmark configuration. The chemical desorption efficiency is evaluated using the formula from \citet{2016A&A...585A..24M}, except for species $O + H$ and $OH + H$ cases, as detailed in Appendix \ref{app: benchmark_network}.

Figure \ref{fig: Nautilus_comparison} compares the temporal evolution of the surface abundances predicted by Nautilus and DESTINY in the single-BE limit for H$_2$O, CO, CH$_4$, NH$_3$, NO, and HCN. The comparison is limited to the pre-monolayer regime based on the Nautilus first layer saturation time ($\sim$ 3500 years). 

\begin{figure}
    \centering
    \includegraphics[width=0.9  \linewidth]{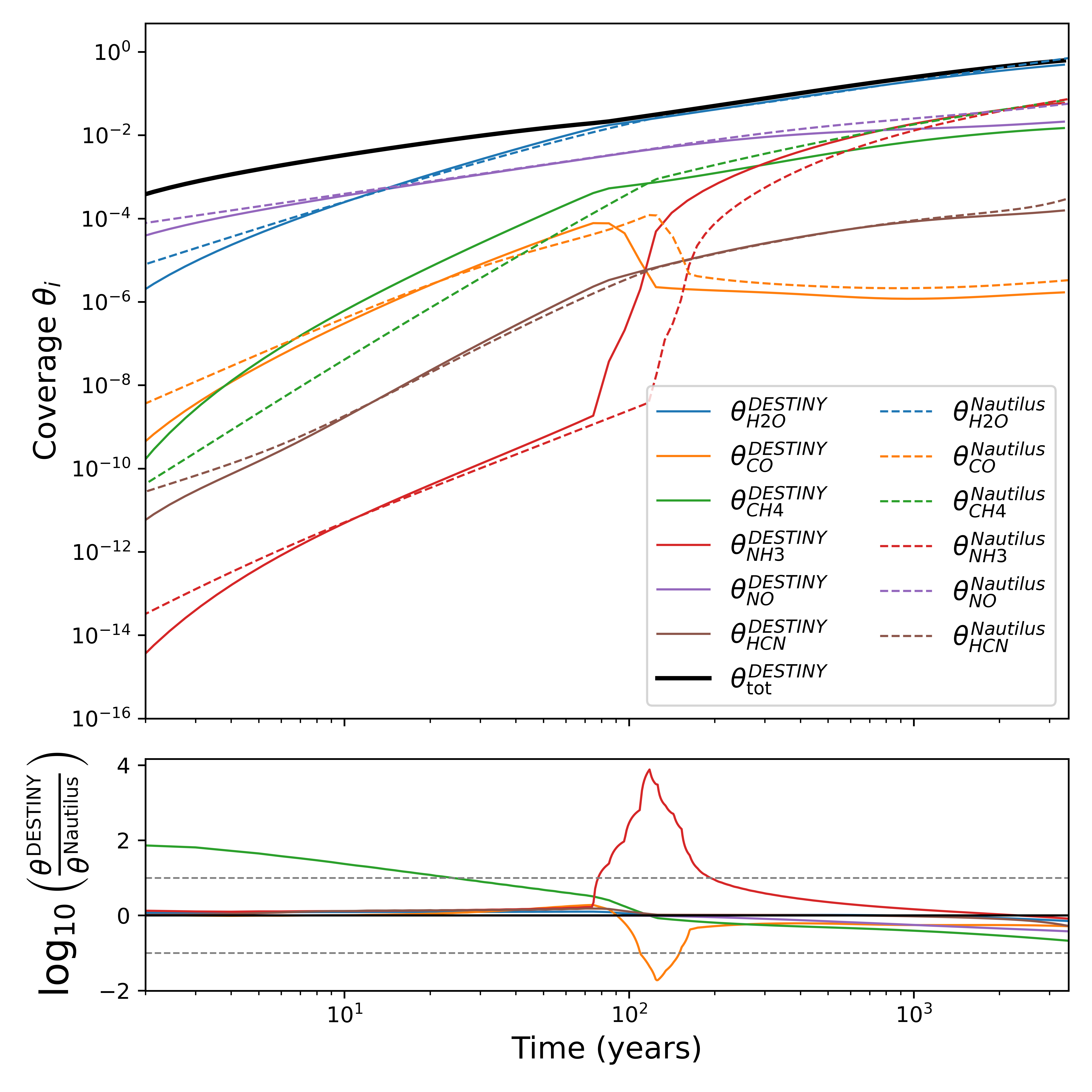}
    \caption{Comparison between the surface coverage predicted by DESTINY and Nautilus in the single-BE limit. 
    Top panel: Temporal evolution of selected surface species coverages, $\theta_i$, together with the total surface coverage, $\theta_{\rm tot}$. Solid lines correspond to DESTINY, while dashed lines show the corresponding Nautilus predictions.
    Bottom panel: Logarithmic abundance ratio between DESTINY and Nautilus, $\log_{10}(\theta_i^{\rm DESTINY}/\theta_i^{\rm Nautilus})$, for the same set of species. The horizontal dashed lines indicate deviations of one order of magnitude. }
    \label{fig: Nautilus_comparison}
\end{figure}

In the case of H$_2$O, NO, and HCN, sticking matches are
observed. CH$_4$ exhibits the largest deviations among the species displayed in Fig. \ref{fig: Nautilus_comparison}, particularly at early times. In this regime, the total surface coverage remains low and the monolayer correction factor $(1-\Theta)$ is essentially inactive. The observed differences therefore primarily reflect the impact of the TFC/Markov-chain probabilistic reformulation itself, free from any influence of the monolayer constraint. Transient deviations are also observed for NH$_3$, although these emerge at later times and are confined within the earlier onset of its sharp abundance increase in DESTINY. Yet, the $(1-\Theta)$ is still not dominant. These deviations are attributable to the inherently different coupling of competitive fluxes and, more specifically, the treatment of the H$_2$ encounter effect; while DESTINY couples the H$_2$ encounter effect with both the diffusivity and desorption efficiency, Nautilus treats it as an enhanced H$_2$ desorption channel. 

The C $\, + \,$ H$_2$ $\rightarrow$ CH$_2$ reaction and subsequent CH$_x$ chain towards CH$_4$ constitute a particularly sensitive diagnostic of the surface H$_2$ chemistry. In contrast to most hydrogenation pathways present in the reduced network, the initiation efficiency of the CH$_x$ chain from atomic C is directly controlled by the availability and mobility of H$_2$ rather than atomic H. This distinction is especially important at early times, where the abundances of C, O, and N (the most abundant species after H$_2$ and He in the initial cloud composition) are primarily governed by accretion from the gas phase. While O and N must first undergo intermediate hydrogenation steps before entering H$_2$-driven pathways (Appendix \ref{app: benchmark_network}), carbon can react directly with H$_2$ through C + H$_2$ $\rightarrow$ CH$_2$. The resulting CH$_2$ can then be further converted into CH$_3$ through CH$_2$ $\, + \,$ H$_2$ $\rightarrow$ CH$_3$ + H before the chemistry becomes increasingly dominated by atomic-H additions.  

The abundances of CH$_2$, CH$_3$, and ultimately CH$_4$ consequently provide a direct probe of the treatment of H$_2$ encounter effect, as the enhanced H$_2$ mobility predicted by DESTINY naturally promotes the H$_2$-driven CH$_4$ formation pathway relative to Nautilus. Since this reaction sequence is one of the few pathways in the reduced network whose initiation is directly controlled by H$_2$, it offers a particularly clean tracer of the impact of the H$_2$ encounter formalism independently of the broader atomic-H chemistry. Additional evidence supporting the role of the different H$_2$ encounter treatments in the CH$_4$ coverage discrepancies is provided in Appendix \ref{app: Naut_comp}. It is also worth mentioning that, in the case of the CH$_x$ chain, the initial atomic carbon is strongly kinetically trapped within its adsorption sites, with a BE of 10.000 $K$. When converted into CH$_2$ by encounter with the diffusive H$_2$, the BE decreases to 1400 $K$. This decreasing trend from the atomic C to the hydrogenated forms CH$_x$ is not retrieved in the case of the NH$_x$ and OH$_x$ chains (Table \ref{tab: BEDs} in Appendix \ref{app: benchmark_network}). This has a positive feedback effect on the enhanced H$_2$ mobility on the CH$_x$ chain abundances towards CH$_4$.

\begin{figure}
    \centering
    \includegraphics[width=0.9  \linewidth]{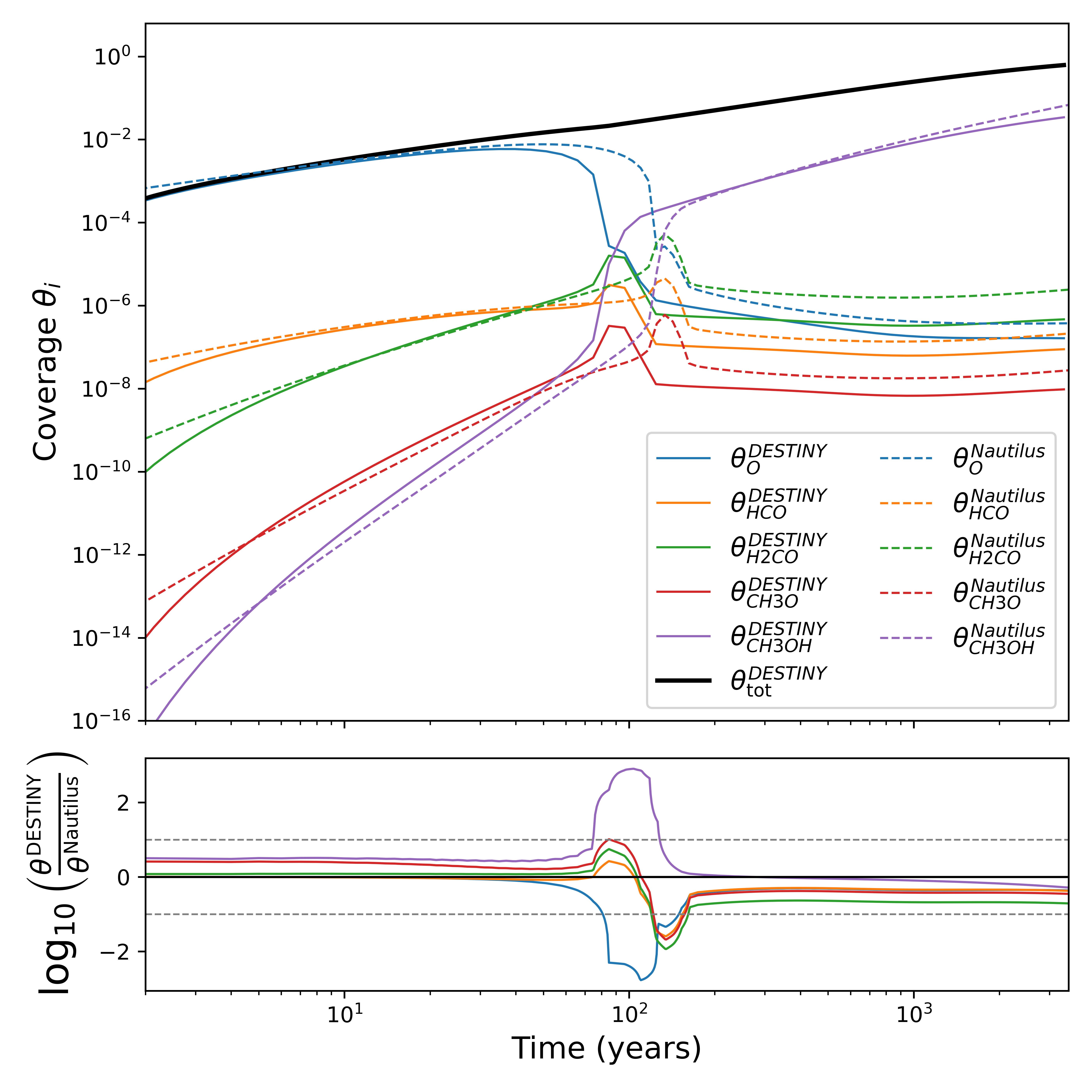}
    \caption{Comparison between the surface coverage predicted by DESTINY and Nautilus in the single-BE limit for HCO- and methanol-related species.  }
    \label{fig: Nautilus_comparison_methanol_and_formaldehyde}
\end{figure}

The transient discrepancy observed in the case of CO, with a lower coverage in the case of DESTINY, can be attributed to the HCO chemistry through the H + CO channel. In this scope, Fig \ref{fig: Nautilus_comparison_methanol_and_formaldehyde} provides the comparison between the pre-monolayer temporal evolution of the surface abundances predicted by Nautilus and DESTINY for HCO and methanol-related species. Actually, the CO discrepancies between DESTINY and Nautilus arise from two successive mechanisms. Around 100 years, they are governed by the earlier rise in the surface abundance of atomic H in DESTINY (Appendix \ref{app: Naut_comp}). As a result, the CO hydrogenation sequence is initiated sooner through CO + H $\rightarrow$ HCO, leading to an earlier enhancement of the HCO abundance. This increase subsequently propagates to H$_2$CO through the barrierless reaction HCO + H $\rightarrow$ H$_2$CO and, in turn, to CH$_3$O and CH$_3$OH through H + H$_2$CO and H + C$_3$O, respectively. In contrast, the onset of the sharp increase in the H surface abundance is delayed in Nautilus. The dominant source of discrepancy shifts to the treatment of the competing H + H$_2$CO reaction channels. As verified from the Nautilus source code and the related list of channel-specific branching ratios, the three reactions implying H + H$_2$CO (Appendix \ref{app: benchmark_network}) are treated independently, meaning that the reaction flux is not partitioned among the competing channels through channel-specific branching ratios. Consequently, each channel consumes H$_2$CO independently, such that the total H$_2$CO destruction rate is given by the sum of the three individual reaction rates. In practice, this is equivalent to counting the same H + H$_2$CO encounter multiple times, leading to an artificial overconsumption of H$_2$CO together with the simultaneous overproduction of CH$_2$OH, CH$_3$O, and HCO. In DESTINY, an explicit channel-specific branching ratios is computed for each reactant pairs harboring different set of products from the ratio of barrier-crossing absolute probabilities, as discussed in Appendix \ref{app: BR}.

At first sight, the independent treatment of the three competing H + H$_2$CO reaction channels in Nautilus should lead to a net depletion of H$_2$CO due to its artificially enhanced destruction. Surprisingly, the opposite is observed. During the early stages, H$_2$CO remains too scarce for this enhanced destruction to significantly affect the HCO abundance. Once H$_2$CO becomes comparable in abundance to HCO, a self-sustaining feedback loop is established. The reaction H + H$_2$CO $\rightarrow$ H$_2$ + HCO overproduces HCO, which is efficiently recycled back into H$_2$CO through the barrierless reaction HCO + H $\rightarrow$ H$_2$CO. The regenerated H$_2$CO is then again overconsumed through the three competing H + H$_2$CO channels, producing additional HCO that is subsequently converted back into H$_2$CO. Consequently, the HCO/H$_2$CO cycle continuously reinforces itself. Owing to the absence of an activation barrier for HCO + H $\rightarrow$ H$_2$CO, the regeneration of H$_2$CO ultimately outweighs its artificial destruction, resulting in the net overproduction of H$_2$CO observed in Nautilus.

While the CH$_4$ discrepancies can be traced back to the different treatment of the H$_2$ encounter effect, DESTINY and Nautilus remain intrinsically different frameworks. Nevertheless, the overall agreement obtained for the remaining species indicates that DESTINY preserves the main chemical trends while introducing physically motivated modifications to the competitive surface chemistry whose effects can be isolated. It should be emphasized that the present benchmark relies on a reduced chemical network specifically designed to isolate the impact of the TFC-based formulation. Consequently, the quantitative differences reported here should not be directly extrapolated to full-scale astrochemical networks, for which the overall impact of the framework remains to be established. Nevertheless, the non-negligible deviations obtained for key species such as CH$_4$ and primary reaction intermediates (Appendix \ref{app: Naut_comp}), especially at times when the factor $1 - \Theta$ negligibly impacts the accretion regime, suggest that measurable consequences can be expected for the formation pathways of more complex molecules. Interestingly, CH$_4$ has previously been identified as one of the key species exhibiting the strongest model-dependent variations across both deterministic and stochastic grain-surface chemistry frameworks \citep{2025A&A...695A.247J}. Given its pronounced sensitivity to the adopted formalism, its behavior in large-scale networks should be revisited to determine whether the TFC-based framework and related H$_2$ encounter effect treatment can alleviate the systematic underproduction of CH$_4$ abundances relative to observations highlighted in the comprehensive benchmark study performed in \citet{2025A&A...695A.247J}. The dust temperature of 12 K was chosen to match the lower-temperature case investigated in \citet{2025A&A...695A.247J}, thereby strengthening the comparison. In the scope of this framework presentation paper, a single representative temperature was considered to keep the discussion focused on the qualitative behavior of DESTINY rather than on its temperature dependence. Nevertheless, since 12 K is commonly regarded as the threshold above which the surface diffusion of O and OH becomes efficient \citep{Angele_Chem_Des}, we repeated the DESTINY--Nautilus comparison at $T_{\rm gas}=T_{\rm dust}=10$ K to assess the robustness of the conclusions presented above. We found that all qualitative trends remain unchanged.

The abundance discrepancies observed for CH$_4$ should also be interpreted in the context of its binding behavior. In \citet{2025A&A...698A.284G}, a larger mean BE was inferred for CH$_4$ (1388 K) than the single value adopted from \cite{2017MolAs...6...22W}. Under the single-BE approximation employed in the present benchmark, this difference has only a limited influence at the considered temperature. However, it becomes particularly relevant when resolving  BEDs since adopting a lower reference BE artificially overemphasizes weakly bound configurations and their associated desorption channels, while lowering the kinetic trapping effect for the stronger BSs. For this reason, the larger mean BE derived in \citet{2025A&A...698A.284G} will be used in the following section when investigating the BED impacts.

\subsection{Effect of the BED-based species coverage discretization}\label{subsec: result_discretiz}

\begin{figure}
    \centering
    \includegraphics[width=0.9  \linewidth]{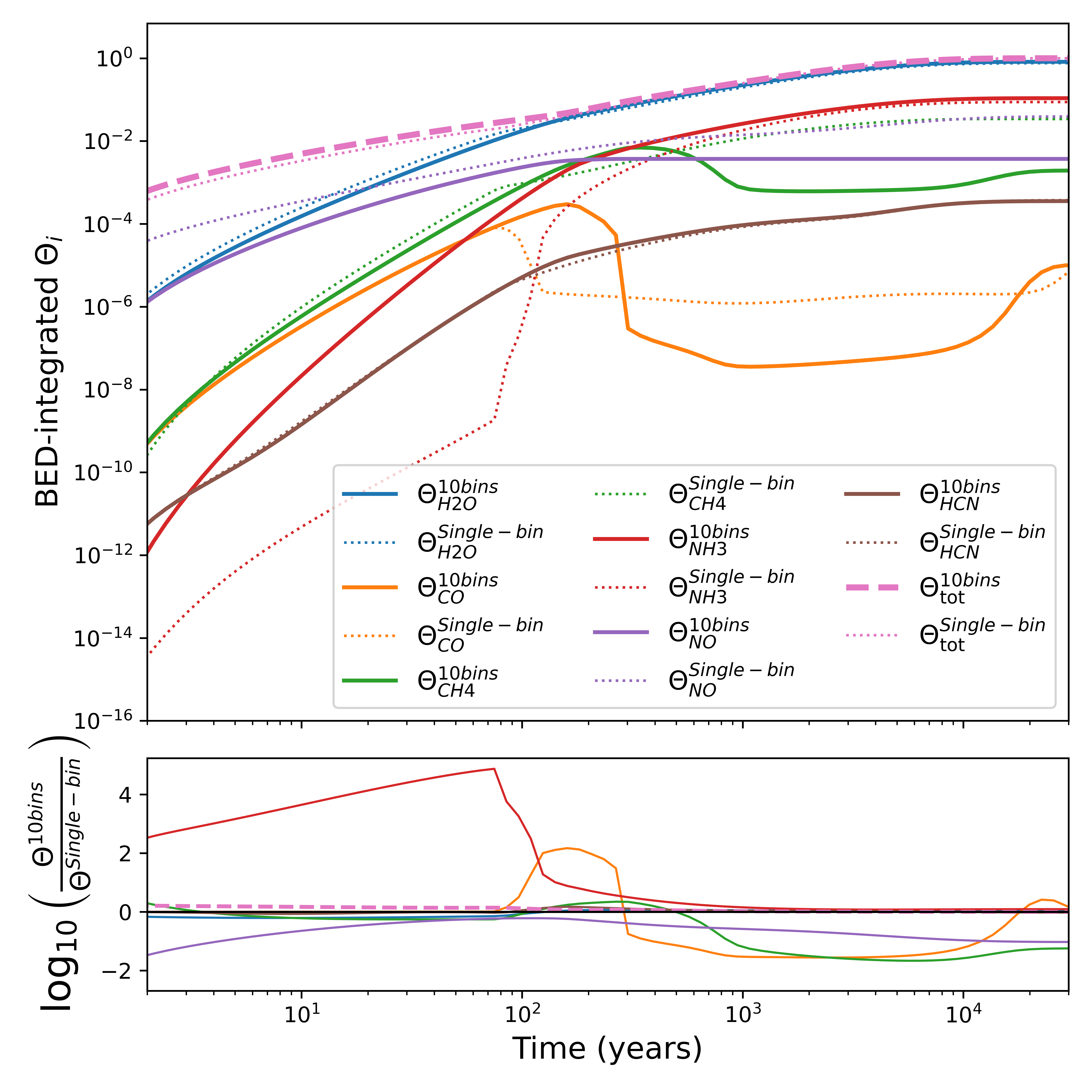}
    \caption{Surface coverages predicted by DESTINY with 10 bins as compared to the single-BE limit. 
    Top panel: Temporal evolution of selected surface species coverages, $\theta_i$, together with the total surface coverage, $\theta_{\rm tot}$. Solid lines correspond to DESTINY, while dashed lines show the corresponding results in the single bin limit.
    Bottom panel: Logarithmic abundance ratio between DESTINY and Nautilus, $\log_{10}(\theta_i^{\rm 10\ bins}/\theta_i^{\rm Single-bin})$, for the same set of species.  }
    \label{fig: BED_comparison}
\end{figure}

A preliminary qualitative assessment of the impact of BED inclusion was conducted by comparing a simulation in the single-bin limit against a multi-bin simulation with a common number of 10 bins per solid-phase species. The underlying physical parameters remain invariant, except that the ER mechanism is activated. Since the current iteration of DESTINY is constrained to the monolayer approximation, physical interpretations are confined to the epoch preceding total monolayer saturation, as the underlying model formalism loses physical validity beyond this threshold, which is reached slightly before $10^4$ years within the current input configuration. As illustrated in Fig. \ref{fig: BED_comparison}, the BED inclusion alters the abundance of critical species, such as NH$_3$, by several orders of magnitude while preserving a nearly identical evolutionary profile for others, such as H$_2$O. These divergent behaviors are elucidated via a comprehensive microscopic pathway analysis, with further insights given in Appendix  \ref{app: multi_bin}.

The most prominent divergence manifests in the evolutionary profile of NH$_3$, whose formation is substantially accelerated from the onset of the simulation, where the monolayer correction term $(1 - \Theta)$ remains negligible. The chemical synthesis pathway towards NH$_3$ is initiated with the sequential atomic hydrogenation of N atoms. In a single-value framework, both atomic precursors possess relatively low BEs ($650\text{ K}$ and $720\text{ K}$ for H and N, respectively). The structural inclusion of a realistic BED introduces a fraction of highly bound BS. This significantly extends the surface residence times of these volatile species, thereby enhancing the probability of reactive encounters with H and H$_2$ (once NH$_2$ is formed), associated with efficient tunneling-assisted diffusivity, especially from the shallower BSs inherent to the inclusion of the BEDs. This subsequently drives the downstream evolution of the NH$_x$ hydrogenation chain. 

Conversely, this behavior does not translate into an increase in the integrated diffusivity, owing to the kinetic competition between trapping mechanisms and rapid diffusion out of shallow BS. This phenomenon is clearly illustrated by the unperturbed evolutionary profile of H$_2$O. Given the higher BE of atomic O ($1600\text{ K}$) with respect to atomic N, it is predominantly trapped within its BS; its characteristic surface residence time is thereby not significantly modified by the distribution at the considered $T_{\rm dust}$, as the characteristic encounter time with highly diffusive and reactive H and H$_2$ partners is smaller than the characteristic inter-bin redistribution and desorption time for atomic O. Any structural enhancement in H$_2$O production would therefore require a net increase in the integrated diffusivity of either atomic H or O, which is not observed within the numerical framework. Similarly, the $\text{HCN}$ abundance profile exhibits a comparable insensitivity to the inclusion of a BED. Within the reduced chemical network, the synthesis of $\text{HCN}$ is exclusively governed by the encounter between highly diffusive atomic $\text{H}$ and the $\text{CN}$ partner. Since the latter possesses an inherently high BE ($2800\text{ K}$), it remains heavily trapped on the grain surface across both model configurations at the considered $T_{\rm dust}$. This already-efficient trapping effectively neutralizes any potential kinetic variations between the single-bin and multi-bin regimes, locking the evolutionary trajectory of $\text{HCN}$ into a BED-independent profile.

The $\text{NO}$ relative underproduction within the multi-bin predictions can be attributed to two primary factors. First, the production rate during the early stages of the simulation is substantially suppressed upon the BED inclusion; this trend is arguably driven by enhanced chemical competition for available atomic N via the LH mechanism in favors of the accelerated $\text{NH}_x$ pathway. Furthermore, the prolonged surface residence times of $\text{N}$ alter the relative weights of competing microchemical pathways; a small shift in importance toward the ER mechanism is observed in the multi-bin simulation (Appendix \ref{app: multi_bin}). Second, enhanced thermal desorption plays a critical role in the BED effect on the NO surface chemistry. In contrast to less abundant species, NO molecules sufficiently accumulate to saturate the deepest potential wells of their BED, leading to the subsequent occupancy of shallower BSs; this configuration effectively elevates the integrated thermal desorption rate for NO. These shallow configurations are continuously replenished by ongoing chemical synthesis via the localized N + O reaction, creating a persistent desorption channel that systematically diminishes the net surface abundance of NO. This physical mechanism remains valid due to both the absence of efficient chemical consumption pathways—for which the characteristic encounter timescale would otherwise be shorter than the desorption timescale—and the relatively low NO mean BE. Consequently, this does not manifest in the case of more heavily bound species, such as OH, H$_2$O, and the NH$_x$ family members, for which thermal desorption remains inefficient across all available BS configurations. 

\begin{figure}
    \centering
    \includegraphics[width=0.9  \linewidth]{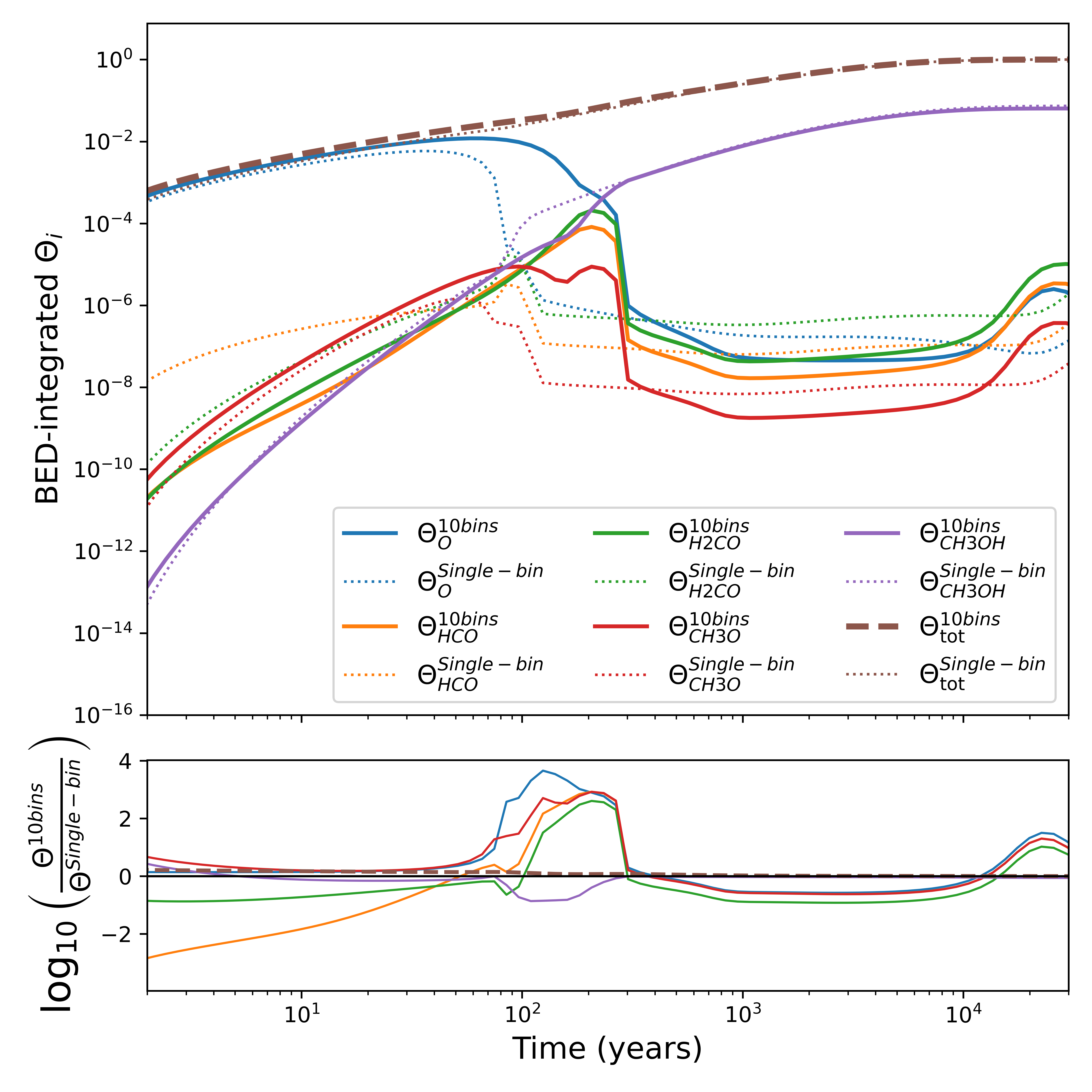}
    \caption{Surface coverages predicted by DESTINY with 10 bins as compared to the single-BE limit for HCO- and methanol-related species. }
    \label{fig: BED_comparison_CHxOx}
\end{figure}

Regarding CH$_4$, it is predominantly regulated by H$_2$-driven chemistry, as previously established. The H$_2$ surface abundance is markedly elevated in the multi-bin simulation (as shown in Appendix \ref{app: multi_bin}), further amplifying the H$_2$ encounter mechanism. However, this kinetic enhancement does not translate into a corresponding increase in the net CH$_4$ fractional coverage. Owing to its intermediate BE (1388 K), the variations induced by the BED introduction are primarily driven by the emergence of shallower BSs, which systematically accelerate thermal desorption. Ultimately, these counteracting physical mechanisms exert opposing kinetic influences and effectively offset one another, leaving the global surface coverage of CH$_4$ fundamentally unaltered.

The CO profile is intrinsically linked to the H$_x$CO and CH$_x$OH$_x$ families, as shown in Fig. \ref{fig: BED_comparison_CHxOx}. During the early stages of ice formation, the HCO and H$_2$CO surface coverages exhibit pronounced discrepancies relative to the single-BE limit; conversely, the abundance profiles of CH$_3$O and CH$_3$OH remain fundamentally unperturbed. These divergent kinetic behaviors are governed by the primary chemical pathways mediating the synthesis of each species and their respective coupling with H$_2$-driven chemistry. Across both the single-bin and multi-bin regimes, methanol is almost exclusively produced via the direct hydrogenation of CH$_3$O. During the initial epoch, this radical precursor is generated via the oxygenation of CH$_3$ mediated by the ER mechanism. Analogous to the trend identified for CH$_4$, the H$_2$-driven intermediate abundances remain largely invariant provided a kinetic equilibrium is established between the local inflation of surface H$_2$ and the concomitant acceleration of thermal desorption—a balancing effect that is manifested for CH$_3$.

Prior to 100 years, HCO is synthesized within the single-bin limit via the addition of atomic O to CH radicals, subsequently initiating the production of H$_2$CO through a sequential addition of H. Conversely, within the multi-bin regime, CH is directly consumed to form CH$_2$ due to the elevated H abundance during the earliest phases of the simulation, thereby effectively shutting down this primary reaction channel. The net abundance of HCO experiences a more pronounced suppression than that of H$_2$CO under multi-bin conditions, owing to a critical shift in the dominant synthetic pathway of the latter. While the direct hydrogenation of HCO is deactivated, an alternative formation route, namely, the oxygenation of CH$_2$ (itself a key product of H$_2$-driven chemistry) becomes dominant during this initial epoch, ultimately mitigating the divergence between the two models.

Beyond 100 years, the HCO, H$_2$CO, and CH$_3$O abundances become noticeably enhanced in the multi-bin configuration compared to the single-BE limit as their synthesis shifts toward the sequential hydrogenation of CO, which has accumulated sufficiently on the grain surfaces. Conversely, the formation of methanol exhibits a minor delay during this intermediate stage of the simulation; this temporary lag is directly correlated with a transient local H depletion, inhibiting the CH$_3$O hydrogenation.

\subsection{Astrochemical implications}

Although the present benchmark relies on a reduced chemical network, the identified mechanisms are expected to have non-negligible consequences in larger astrochemical systems.

The treatment of the H$_2$ encounter effect has been shown to modify the efficiency of H$_2$-driven surface chemistry. The largest deviations are found within the CH$_x$ hydrogenation sequence, where the impact of the probabilistic formulation can be isolated from monolayer effects during the early phase. Since CH$_4$ has previously been identified as one of the most model-sensitive grain-surface species \citep{2025A&A...695A.247J}, these results suggest that the treatment of H$_2$ surface exploration may constitute a measurable source of uncertainty in current astrochemical models. Regarding the effect of BEDs, their introduction alters the balance between diffusion, desorption, and reaction by redistributing the surface population among BSs of different depths. This is expected to propagate throughout larger astrochemical networks.

\section{Conclusions and outlooks}\label{sec: ccl_perspect}

In this work, we introduced DESTINY, a deterministic solid-phase astrochemical framework based on a TFC formalism under an absorbing Markov-chain representation of grain-surface competitive processes. By reformulating diffusion, desorption, and reactive encounters as competing Markov states, together with idle states, the framework derives effective event probabilities from the stationary distribution of the corresponding Markov chain while preserving a deterministic ODE description. This formalism is resolved onto the binding-energy distributions (BEDs) of surface species, enabling the explicit treatment of energetic heterogeneity across amorphous interstellar ice surfaces.

Benchmark comparisons against the widely used Nautilus code have shown that the probabilistic reformulation alone can produce measurable differences in the predicted surface abundances, even in the single-BE limit. This is especially true for the CH$_x$ chain, for which the impact of the DESTINY framework on the treatment of the H$_2$ encounter effect, with an enhanced H$_2$ residence time and mobility, and the resulting modulation of the H$_2$-driven chemistry can be isolated based on the reduced surface network used in this work. The same comment holds for the NH$_x$ chain, at least in a multi-bin configuration. Furthermore, the BED-resolved formulation naturally provides access to quantities that remain inaccessible to conventional single-BE approaches, including BE-resolved occupancies, process-specific contributions, and the explicit coupling between surface chemistry and the energetic landscape of amorphous ice. Reported deviations from the single-BE limit are expected to have a measurable influence on the chemistry of a large-scale surface network. 

Several developments naturally emerge from the present work. First, the framework should be extended to include non-diffusive chemistry mechanisms \citep{non_diff_chem_GH20,2022ApJS..259....1G}. Second, systematic sensitivity analyses should be conducted on both BED parameters and key physical quantities known to strongly influence surface chemistry, including species-dependent diffusion-to-desorption ratios $\chi$, diffusion and reaction barrier widths, chemical desorption parameters, ... Particular attention should also be devoted to the H$_2$ encounter effect, whose treatment has been shown to significantly affect the predicted abundances of several key species. Future investigations should therefore assess the sensitivity of the model to the adopted H$_2$ encounter binding energy and associated $\chi$ value, as well as explore the inclusion of analogous effects for atomic H on H$_2$-rich substrates \citep{10.3389/fspas.2021.671622,Chang_2021}. This would provide new insights for the hunt for sources of discrepancies between simple RE-based model outputs and actual ice observations in the case of H and H$_2$ driven-chemistry (e.g., CH$_4$ and NH$_3$ as highlighted from the reduced network used in the work). Finally, the extension of DESTINY toward multilayer ice mantles and its application to large-scale astrochemical networks, including complex organic molecules, constitute the next major steps toward its comprehensive BED-resolved description of interstellar grain-surface chemistry.

\begin{acknowledgements}
    The authors thank the referee, Dr. Angèle Taillard, for her time as well as for the detailed and positive report. Her thoughtful comments prompted some refinements that improved the clarity and depth of the discussions of the results. 
    M. Groyne and C. Baijot are Research Fellows of the Belgian National Fund For Scientific Research (F.R.S.-FNRS). 
\end{acknowledgements}

\bibliographystyle{aa} 
\bibliography{ref} 

\begin{appendix}
\nolinenumbers 

\section{Encounter-dependent BEs} \label{app: Encounter BE}

\subsection{Effect of the encounter-dependent BEs on BE-resolved state counts}

Currently, the only encounter-dependent binding energy implemented within the DESTINY framework corresponds to H$_2$-H$_2$ interactions. However, the following encounter formalism remains strictly generalized to facilitate the integration of additional encounter mechanisms in future investigations. When a migratory species $i$ diffuses onto a binding site of energy $E_i$ already occupied by a species $j$ with a binding energy $E_j$, their respective binding energies are potentially modified via mutual interactions to the values $q_{i, j}(E_i)$ and $q_{j, i}(E_j)$. Here, the operator $q_{i, j}$ represents a perturbative function. For the specific case of the H$_2$ encounter mechanism, this is evaluated as $q_{H_{2}, H_{2}}(E_{H_2}) = 23$~K. This local modification of the binding energies directly impacts the relative probabilities and trial frequencies used to evaluate the counts of transient states, as formulated in the following equations.

\begin{align}
    \notag
    \#&_{x}^{transient/terminal} (E_i) = \frac{P_{\rm diff, \rm gateway}^{eff, i}}{(1 - P_{\rm survival}^{\rm transient, eff, i})} \Bigg[ \\ \notag
     & \frac{1}{\nu_{i, 0}(E_i)}(1 - \Theta) \frac{(1 - \theta_i (E_i))}{(1 - \Theta_i )} P_{x}^{rel, mono} (E_i)\\ \notag
     & + \sum_{j \neq i} \frac{1}{\nu_{i, 0}(q_{i, j}(E_i))} \Bigg( \int \theta_j (E_j)g_j (E_j) \frac{(1 - \theta_i (E_i))}{(1 - \Theta_i)} (1 - s_{\rm barrierless}^{i+j}) \times \\ \notag
     &  \biggl[P_{x}^{rel, double} (q_{i, j}(E_i), q_{j, i}(E_j)) \\\notag
     &+ (P_{diff}^{rel, double}(q_{j, i}(E_j), q_{i, j}(E_i)) + P_{des}^{rel, double} (q_{j, i}(E_j), q_{i, j}(E_i))) \\ \notag 
     & \times P_{x}^{rel, mono}(E_i)\biggl] dE_j \Bigg) \\ \notag
     & + \frac{1}{\nu_{i, 0}(q_{i, i}(E_i))} \theta_i (E_i) (1 - s_{\rm barrierless}^{i+i}) \times \biggl[P_{x}^{rel, double} (q_{i, i}(E_i), q_{i, i}(E_i)) 
     \\\notag 
     &+ (P_{diff}^{rel, double} (q_{i, i}(E_i), q_{i, i}(E_i)) + P_{des}^{rel, double} (q_{i, i}(E_i), q_{i, i}(E_i))) \\ \notag 
     & \times P_{x}^{rel, mono}(E_i)\biggl] \Bigg]
\end{align}

\begin{align}
    \notag
    \#&_{idle}^{transient/terminal} (E_i) = \frac{P_{\rm diff, \rm gateway}^{eff, i}}{(1 - P_{\rm survival}^{\rm transient,  eff, i})} \Bigg[ \\ \notag
     & \frac{1}{\nu_{i, 0}(E_i)}(1 - \Theta) \frac{(1 - \theta_i (E_i))}{(1 - \Theta_i)} P_{idle}^{rel, mono} (E_i)\\ \notag
     & + \sum_{j \neq i} \frac{1}{\nu_{i, 0}(q_{i, j}(E_i))} \Bigg( \int \theta_j (E_j)g_j (E_j) \frac{(1 - \theta_i (E_i))}{(1 - \Theta_i)} (1 - s_{\rm barrierless}^{i+j}) \times \\ \notag
     &  \biggl[\frac{P_{idle}^{rel, double} (q_{i, j}(E_i), q_{j, i}(E_j))}{P_{evol}^{double} (q_{i, j}(E_i), q_{j, i}(E_j))} \\\notag
     &+ (P_{diff}^{rel, double}(q_{j, i}(E_j), q_{i, j}(E_i)) + P_{des}^{rel, double} (q_{j, i}(E_j), q_{i, j}(E_i))) \\ \notag 
     & \times P_{idle}^{rel, mono}(E_i)\biggl] dE_j \Bigg) \\ \notag
     & + \frac{1}{\nu_{i, 0}(q_{i, i}(E_i))} \theta_i (E_i) (1 - s_{\rm barrierless}^{i+i}) \times \biggl[\frac{P_{idle}^{rel, double} (q_{i, i}(E_i), q_{i, i}(E_i))}{P_{evol}^{double} (q_{i, i}(E_i), q_{i, i}(E_i))} 
     \\\notag 
     &+ (P_{diff}^{rel, double} (q_{i, i}(E_i), q_{i, i}(E_i)) + P_{des}^{rel, double} (q_{i, i}(E_i), q_{i, i}(E_i))) \\ \notag 
     & \times P_{idle}^{rel, mono}(E_i)\biggl] \Bigg]
\end{align}

\noindent where $x$ indicates "diffusion" or "desorption" probabilities. The perturbative function is directly integrated into the evaluation of reactive counts by modifying the underlying binding energies utilized within the reaction probability formulations of Eq. \ref{eq: coutingevent_tot_reac_Ei}.

\subsection{Effect of the encounter-dependent BEs on effective trial frequencies for net event rates}

Because macroscopic events encompass multiple distinct microscopic pathways, each characterized by distinct trial frequencies, the frequency terms used in the derivation of the rate equations (Eq. \ref{eq: RateEvent_i_E_i}) must be replaced by generalized effective frequencies. These effective frequencies are explicitly evaluated by averaging the local frequency (i.e., $\nu_{i, 0}(q_{i, j}(E_i))$ or $\nu_{i,j}(q_{i, j} (E_i), q_{j, i} (E_j))$ for reactive encounters) weighted over the discrete contributions of each active microscopic pathway. The formulation of an effective frequency is inherently required for reactive pathways even in the absence of perturbative functions. As the underlying trial frequency for reactions (Eq. \ref{eq: trial_frequency_reac}) fundamentally depends on the binding energy distribution of the partner species $j$, the implementation of a generalized effective trial frequency $\nu_{i, j}(q_{i, j} (E_i), q_{j, i} (E_j))$ remains mathematically mandatory. Both formulations are defined explicitly by the following expressions:

\begin{align}
    \notag
    \nu_{i, 0}^{x} (E_i) &= \Bigg \{ \theta_i (E_i) P_{x}^{rel, mono} (E_i) \\ \notag
    & + \frac{P_{\rm diff, \rm gateway}^{eff, i}}{(1 - P_{\rm survival}^{\rm transient, eff, i})} \Bigg[ \\ \notag
     & (1 - \Theta) \frac{(1 - \theta_i (E_i))}{(1 - \Theta_i)} P_{x}^{rel, mono} (E_i)\\ \notag
     & + \sum_{j \neq i}  \Bigg( \int \theta_j (E_j)g_j (E_j) \frac{(1 - \theta_i (E_i))}{(1 - \Theta_i)} (1 - s_{\rm barrierless}^{i+j}) \times \\ \notag
     &  \biggl[P_{x}^{rel, double} (q_{i, j}(E_i), q_{j, i}(E_j)) \\\notag
     &+ (P_{diff}^{rel, double}(q_{j, i}(E_j), q_{i, j}(E_i)) + P_{des}^{rel, double} (q_{j, i}(E_j), q_{i, j}(E_i))) \\ \notag 
     & \times P_{x}^{rel, mono}(E_i)\biggl] dE_j \Bigg) \\ \notag
     & + \theta_i (E_i) (1 - s_{\rm barrierless}^{i+i}) \times \biggl[P_{x}^{rel, double} (q_{i, i}(E_i), q_{i, i}(E_i)) 
     \\\notag 
     &+ (P_{diff}^{rel, double} (q_{i, i}(E_i), q_{i, i}(E_i)) + P_{des}^{rel, double} (q_{i, i}(E_i), q_{i, i}(E_i))) \\ \notag 
     & \times P_{x}^{rel, mono}(E_i)\biggl] \Bigg] \Bigg \} / \{ \#_{x}^{start} (E_i) +  \#_{x}^{transient/terminal} (E_i) \}
\end{align}

\begin{align} \label{eq: eff_trial_frequency_reac}
    \nu_{i, j} (E_i) &= \Bigg ( \int \theta_j (E_j)g_j (E_j) \frac{(1 - \theta_i (E_i))}{(1 - \Theta_i)} \times \notag \\ \notag
     & \{ s_{\rm barrierless}^{i+j} + (1 - s_{\rm barrierless}^{i+j}) P_{reac, i+j}^{rel, double} (q_{i, j} (E_i), q_{j, i} (E_j)) \} \, dE_j \Bigg ) \\ \notag
     &  / \Bigg ( \int \frac{1}{\nu_{i,j}(q_{i, j} (E_i),q_{j, i} (E_j))} \theta_j (E_j)g_j (E_j) \frac{(1 - \theta_i (E_i))}{(1 - \Theta_i)} \times \\ \notag
     & \{ s_{\rm barrierless}^{i+j} + (1 - s_{\rm barrierless}^{i+j}) P_{reac, i+j}^{rel, double} (q_{i, j} (E_i), q_{j, i} (E_j)) \} \, dE_j \Bigg ) \notag
\end{align}

\newpage

\section{Branching ratio for detailed TFC contributions}\label{app: BR}

The absolute event probabilities introduced throughout the main text combine several microscopic pathways contributing to the same macroscopic event. In order to recover the individual contributions of these pathways when constructing the corresponding gain and loss rates, branching ratios are introduced. They quantify the conditional probability that a given event originates from a specific microscopic channel, knowing that the event itself has occurred.

We first consider the desorption process. As discussed in Sect. \ref{sec: tfc-formalism}, desorption results from two mutually exclusive thermal states, namely the quiescent grain temperature $T_{\rm dust}^{quiescent}$ and the transient cosmic-ray-heated temperature $T_{\rm peak}^{\rm CR-heating}$. The corresponding branching ratios are

\begin{align}
    &BR_{des}^{T_{\rm peak}^{\rm CR-heating}}(E_i)
    =
    \frac{
    P_{des}^{CR-heating}(E_i)
    }{
    P_{des}(E_i)
    }
    \\
    &BR_{des}^{T_{\rm dust}^{quiescent}}(E_i)
    =
    \frac{
    \left(1-f_{\rm peak}^{CR-heating}\right)
    P_{des}^{T_{\rm dust}^{quiescent}}(E_i)
    }{
    P_{des}(E_i)
    }
\end{align}

\noindent such that

\begin{equation}
    \sum BR_{des}^{T}(E_i)
    =
    1.
\end{equation}

For diffusion, the thermally promoted and tunneling-mediated pathways are treated as independent contributions to the same diffusive Markov state. Analogously to the competitive weighting used in the definition of the relative probabilities, the contribution of each independent diffusion pathway is obtained from its relative weight within the total set of independent diffusion pathways, as 

\begin{align}
&BR_{diff}^{thermal}(E_i)
=
\frac{
P_{diff}^{thermal}(E_i)
}{
P_{diff}^{thermal}(E_i)+P_{diff}^{tun}(E_i)
},
\\
&BR_{diff}^{tun}(E_i)
=
\frac{
P_{diff}^{tun}(E_i)
}{
P_{diff}^{thermal}(E_i)+P_{diff}^{tun}(E_i)
}.
\end{align}

\noindent satisfying

\begin{equation}
BR_{diff}^{thermal}(E_i)+BR_{diff}^{tun}(E_i)=1,
\end{equation}

The detailed sub-contributions for $BR_{diff}^{thermal}(E_i)$ from CR-heated and quiescent thermal state are given by

\begin{align}
    &BR_{diff}^{T_{\rm peak}^{\rm CR-heating}}(E_i)
    =
    \frac{
    P_{diff}^{CR-heating}(E_i)
    }{
    P_{diff}^{thermal}(E_i)
    } BR_{diff}^{thermal}(E_i)
    \\
    &BR_{diff}^{T_{\rm dust}^{quiescent}}(E_i)
    =
    \frac{
    \left(1-f_{\rm peak}^{CR-heating}\right)
    P_{diff}^{T_{\rm dust}^{quiescent}}(E_i)
    }{
    P_{diff}^{thermal}(E_i)
    } BR_{diff}^{thermal}(E_i)
\end{align}

\noindent such that

\begin{equation}
    \sum BR_{diff}^{T}(E_i)
    =
    BR_{diff}^{thermal}(E_i)
\end{equation}

Similar expressions are retrieved in the case of thermally-activated and tunneling mediated reactive contributions. Indeed, 

\begin{align}
&BR_{thermal}^{i, j}
=
\frac{
P_{thermal}^{i, j}
}{
P_{thermal}^{i, j}+P_{tunn}^{i, j}
},
\\
&BR_{tunn}^{i,j}
=
\frac{
P_{tunn}^{i, j}
}{
P_{thermal}^{i, j} + P_{tunn}^{i, j}
}.
\end{align}

\noindent satisfying

\begin{equation}
BR_{thermal}^{i,j}+BR_{tunn}^{i,j}=1,
\end{equation}

The detailed thermally activated reactive contributions are redistributed according to 

\begin{align}
    &BR_{T_{\rm peak}^{\rm CR-heating}}^{i,j}
    =
    \frac{
    P_{CR-heating}^{i,j}
    }{
    P_{thermal}^{i,j}
    } BR_{thermal}^{i,j}
    \\
    &BR_{T_{\rm dust}^{quiescent}}^{i,j}
    =
    \frac{
    \left(1-f_{\rm peak}^{CR-heating}\right)
    P_{T_{\rm dust}^{quiescent}}^{i,j}
    }{
    P_{thermal}^{i,j}
    } BR_{thermal}^{i,j}
\end{align}

\noindent such that

\begin{equation}
    \sum BR_{T}^{i,j}
    =
    BR_{thermal}^{i,j}
\end{equation}

Finally, a branching ratio should be introduced for competitive $i+j$ channels from the same reactive pair. It accounts for the fact that a given reactive pair may lead to several distinct reaction channels associated with different product sets. While the total reaction probability entering the relative probabilities is obtained from $P_{cross,i+j}^{tot,sum}$ (Eq. \ref{eq: Preac_eff_from_double}), the individual product channels must be recovered when constructing the gain terms in order to ensure mass conservation at the ODE level. Unlike the discussed above, these are therefore of absolute importance for the ODE-level information. The branching ratio associated with a reaction channel $c_k$ originating from the reactive pair $(i+j)$ is therefore defined as

\begin{equation}
BR_{cross}^{i+j,c_k} = 
\frac{
P_{cross}^{i+j,c_k}
}{
P_{cross,i+j}^{tot,sum}
},
\end{equation}

\noindent By construction,

\begin{equation}
\sum_{c_k}
BR_{cross}^{i+j,c_k} = 1
\end{equation}

These branching ratios correspond to the factors appearing explicitly in Eq. \ref{eq: R_gain}. They constitute the only branching ratios directly required by the ODE system, as they determine how the total reactive flux associated with a given reactive pair is partitioned among the different product channels.

\section{Dust-to-gas BED-integrated coupling}
\label{app: gas_coupling}

The dust-to-gas coupling loss term is defined in Eq. \ref{eq: R_loss_gas} and encapsulates the contributions of the accretion and the ER reactive pathway. Conversely, the corresponding gain term, formulated in Eq. \ref{eq: R_des_gas}, encompasses both thermal and chemical desorption mechanisms. It is worth highlighting that the chemical desorption dust-to-gas coupling gain contribution should be considered from $f_{cd}^{j + k}$, in opposition to the solid-phase counterpart, corrected by $(1-f_{cd}^{j + k})$. 

\begin{align}
    R_{i, dust \, to \, gas}^{loss} &= (1 - \Theta) S_i \sigma_{gr} v_{th,i} n_i n_{gr} \label{eq: R_loss_gas}\\ \notag
     &+ \sum_j \Theta_j S_i \sigma_{gr} v_{th,i} n_i n_{gr} P_{cross}^{i+j}
\end{align}

\begin{align} \label{eq: R_des_gas}
    R_{i, dust \, to \, gas}^{gain}& = \int R_{des} (E_i') + \frac{(1 - \theta_i (E_i'))}{(1 - \Theta_i)} \\ \notag
     &\times  \sum_{\substack{j + k \rightarrow i + ... \\ j \neq k}} \bigg[f_{cd}^{j + k} (E_i') \alpha_{j + k \rightarrow i}^{stoechio} \frac{P_{cross}^{j+k, c_i}}{P_{cross, j+k}^{tot, sum}} \\ \notag
     & \times \bigg(\int (R_{reac, j + k} (E_j) + R_{reac, j + k}^{ER} (E_j)) \times g_j (E_j) \, dE_j \\\notag
     & + \int (R_{reac, k + j} (E_k) + R_{reac, k + j}^{ER} (E_k)) \times  g_k (E_k) \, dE_k \bigg) \bigg] \\
     \notag\\
     & +
     \frac{(1 - \theta_i (E_i))}{(1 - \Theta_i)} \times \sum_{j + j \rightarrow i + ...}
     \bigg[
     f_{cd}^{j + j} (E_i')
     \alpha_{j + j \rightarrow i}^{stoechio}
     \frac{P_{cross}^{j + j \rightarrow i + ...}}{P_{cross, j+j}^{tot, sum}}
     \notag\\\notag
     & \times     
     \int (R_{reac, j + j} (E_j) + R_{reac, j + j}^{ER} (E_j))
     \times g_j (E_j) \, dE_j
     \bigg]
     \Bigg )
     \\ \notag 
     & \times g_i (E_i') \, dE_i' \times N_{site} \cdot n_{gr} 
\end{align}

\noindent where $n_{gr}$ is the dust grain number density.

The complete expression for the gas-phase ODE system RHS, considering the above dynamical coupling terms with the solid-phase, is given in Eq. \ref{eq: gas_phase_full_RHS}. 

\begin{align}\label{eq: gas_phase_full_RHS}
    \frac{dn_i}{dt} &= \sum_k \sum_l \alpha_{k + l \rightarrow i}^{stoechio} k_{kl} n_k n_l + \sum_k \alpha_{k +(s)UV/CR \rightarrow i}^{stoechio} k_k^{(s)UV/CR} n_k \\ \notag
    &- \alpha_{i + j \rightarrow ...}^{stoechio} n_i \sum_j k_{ij} n_j - k_{i + (s)UV/CR \rightarrow ...} n_i \\ \notag
    & + R_{i, dust \, to \, gas}^{gain} - R_{i, dust \, to \, gas}^{loss}
\end{align}

\section{Desorption contributions through energetic processing}\label{app: ice_proc}

As commented in the main text, in the present formulation of a monolayer, energetic processing is restricted to enhanced desorption. Surface dissociation is purposely omitted from this discussion due to its inherent net gain contribution to the coverage. This would violate the site conservation from the monolayer approximation (Eq. \ref{eq: theta_cdt}) under high-coverage conditions.

The UV-induced desorption formalism of \citet{2016MNRAS.459.3756R} is discretized over the BED, yielding Eqs. \ref{eq: RUV_discretized} and \ref{eq: RUV_secondary_discretized} for interstellar and cosmic-ray-induced UV photons, respectively.

\begin{align} \label{eq: RUV_discretized}
    R_{UV}^{loss} (E_i) &=  \theta_i (E_i) F_{UV} S_{UV} Y_{pd} \frac{\sigma_{gr}}{N_{site}} e^{- 2 A_V}
\end{align}

\begin{align} \label{eq: RUV_secondary_discretized}
    R_{UV-CR}^{loss} (E_i) &= \theta_i (E_i) F_{UV-CR} S_{UV-CR} Y_{pd} \frac{\sigma_{gr}}{N_{site}}
\end{align}

\noindent where $F_{\text{UV}}$ denotes the UV flux in photons cm$^{-2}$ s$^{-1}$, $S_{UV}$ represents the scaling factor scaling the UV radiation field, $A_v$ is the visual extinction parameter, and $Y_{\text{pd}}$ is the photo-desorption yield in molecules photon$^{-1}$. Following the recommendation of \cite{2013ApJ...779..120B}, and justified by the current lack of empirical or theoretical constraints on photo-desorption, the yield parameter $Y_{pd}$ is treated as species-independent. The exponential factor embedded within Eq. \ref{eq: RUV_secondary_discretized} accounts for the attenuation of the UV field as a function of the local extinction $A_v$ \citep{1991ApJS...77..287R}.

Since desorption inherently couples the monolayer system to the gas-phase reservoir, evaluating the corresponding gas-phase gain rates necessitates the formulation of bin-integrated representations for both the primary and secondary UV-desorption equations, which are given, respectively, by Eqs. \ref{eq: RUV_integrated} and \ref{eq: RUV_secondary_integrated}.

\begin{align} \label{eq: RUV_integrated}
    R_{UV}^{gain, dust \, to \, gas} &= \Theta_i n_{gr} \phi_{UV} S_{UV} Y_{pd} \sigma_{gr} e^{- 2 A_V}
\end{align}

\begin{align} \label{eq: RUV_secondary_integrated}
    R_{UV-CR}^{gain, dust \, to \, gas} &= \Theta_i n_{gr} \phi_{UV-CR} S_{UV-CR} Y_{pd} \sigma_{gr}
\end{align}

Similarly, CR-induced sputtering with dust grains is straightforwardly incorporated into the present framework following the prescription of \cite{2021A&A...652A..63W}. Its discretized formulation is explicitly defined in Eq. \ref{eq: R_sputtering_discretized} \citep{2021A&A...652A..63W}, while the corresponding bin-integrated counterpart is given in Eq. \ref{eq: R_sputtering_integrated}.

\begin{align} \label{eq: R_sputtering_discretized}
    R_{CR, sputtering}^{loss} (E_i) = \theta_i (E_i) \times (\frac{\zeta_{CR}}{3 \times 10^{-17}}) \times Y_{eff} \times \frac{\sigma_{gr}}{N_{site}}
\end{align}

\begin{align} \label{eq: R_sputtering_integrated}
    R_{CR, sputtering}^{gain, dust \, to \, gas} (E_i) = \Theta_i n_{gr} \times (\frac{\zeta_{CR}}{3 \times 10^{-17}}) \times Y_{eff} \times \sigma_{gr}
\end{align}

\noindent with

\begin{equation}
    Y_{eff} = Y^{\infty} \times (1 - e^{-(\frac{n_{layers}}{\beta})^{\gamma}}) \notag
\end{equation}

\noindent where $Y_{eff}$ denotes the effective desorption yield integrated over the CR spectral distribution, $Y^{\infty}$ represents the asymptotic sputtering yield characteristic of thick ice substrates, $\beta$ and $\gamma$ are parameters scaling with the ice composition, and $n_{layers}$ is parameterized to $\Theta$ within the context of an evolving monolayer system. $n_{layers}$ allows for the consideration of the CR penetration depth \citep{2018A&A...618A.173D}, which would become revenant in a multi-layer system. Regarding the predominance of water in quiescent clouds considered in this work, water-ice CR-sputtering values from \cite{2018A&A...618A.173D} for $Y^{\infty}$, $\beta$ and $\gamma$ are assumed. It should be emphasized that molecular dissociation is currently not explicitly treated for sputtering-induced desorption events. The impact of such simplifying considerations should be interesting to consider in subsequent studies by simply adding a branching ratio for each (un)fragmented set of desorbed species \citep{2009ApJ...693.1209O, 2014FaDi..168..533F, 2025A&A...693A..30F}. 

\section{Numerical switches} \label{app: Switches}

DESTINY architecture integrates configurable switches designed to selectively activate alternative numerical methods, specialized astrochemical processes, and their corresponding physical parameters. A comprehensive compilation of these input parameters with values used in this work and operational switches is presented in Table \ref{tab: inputs} and Table \ref{tab: switches}, respectively.

\begin{table}[H]
    \centering
    \caption{Input parameters defined in DESTINY.}
    \begin{tabular}{l l l}
    \hline
    \hline
        Input parameter & Abbreviation & Value \\ \hline \hline
        Physical parameters & & \\ \hline \hline
        Mean total hydrogen density & $n_H$ [cm$^{-3}$] & $3 \times 10^4$ \\
        Gas temperature & $T_{gas}$ [K] &  $12$ \\
        Dust temperature & $T_{dust}$ [K] &  $12$ \\
        CR ionization rate & $\zeta_{CR}$ [s$^{-1}$] & $1.3 \times 10^{-17}$ \\
        Visual extinction & $A_v$ & $15$  \\
        UV flux & $F_{UV}$ [G$_0$]\tablefootmark{a} & $1.00$ \\
        Dust to gas mass ratio & $1 \times 10^{-2}$ \\
        Surface site density\tablefootmark{b} & $n_{site}$ [cm$^{-2}$] & $1.5 \times 10^{15}$ \\
        Dust radius & $r_{grain}$ [cm] & $1 \times 10^{-5}$ \\
        Grain density & $\rho_{dust}$ [g/cm$^{-3}$] & $3$ \\
        Reaction barrier width & $a_{reac}$ [\AA] & $1.00$ \\
        Diffusion barrier width & $a_{diff}$ [\AA] & $2.50$ \\
        Photo-desorption yield & $Y_{pd}^{std}$\tablefootmark{c} & $1 \times 10^{-4}$ \\
         & $Y_{pd}^{secp}$\tablefootmark{d} & $1 \times 10^{-4}$ \\
        Sputtering parameters\tablefootmark{e} & $Y^{\infty}$ & $3.63$ \\
        & $\beta$ & $3.25$ \\
         & $\gamma$ & $0.67$ \\
        Chemical desorption factor  & $f_{cd}$\tablefootmark{f} & $1 \times 10^{-2}$ \\
         & $f_{cd}^{multi}$\tablefootmark{g} & $1 \times 10^{-3}$ \\
        CR heating peak duration & $t_{peak}^{CR-heating}$ [s] & $1 \times 10^{-5}$ \\
        CR heating peak temperature & $T_{peak}^{CR-heating}$ [K] & $70$ \\
        Fe ionization rate & $\zeta_{Fe+}^{Grain}$ [s$^{-1}$] & $3 \times 10^{-14}$ \\
        Diffusion-to-desorption ratio & $\chi$ & $0.4$\tablefootmark{h} \\
        Trial frequency & $\nu_{x, 0}$ & $1 \times 10^{12}$ \\
        Number of H$_2$O molecules\tablefootmark{i} & $n_{H_2O}$ & $5$ \\ \hline \hline
        Solver parameters & & \\ \hline \hline
        Start time & $t_0$ [s] & $0$ \\
        End time & $t_f$ [s] & $10^7$ \\
        Absolute $n_i$ error\tablefootmark{j} & NA & $10^{-20}$ \\
        Absolute $\theta_i(E_i)$ error\tablefootmark{j} & NA & $10^{-20}$ \\
        Relative error\tablefootmark{j} & NA & $10^{-6}$ \\
        Number of steps\tablefootmark{k} & NA & $128$ \\ \hline \hline
        BED-related parameters \\ \hline \hline        
        Fixed BE resolution & $\Delta E [K]$ & $150$ \\
        Fixed number of bins & $n_{bins}$ & $10$ \\
        Number of sigmas\tablefootmark{l} & $n_{sigma}$ & $3$ \\

    \hline
    \end{tabular}
    \tablefoot{
    \tablefoottext{a}{In Draine unit.}
    \tablefoottext{b}{The number of site per dust grains ($N_{site}$) is not defined as a user-defined parameter, but is computed from $N_{site} = 4\pi r_{grain}^2 \rho_s$.}
    \tablefoottext{c}{Photo-desorption yield from standard UV photons in molecules photon$^{-1}$.}
    \tablefoottext{d}{Photo-desorption yield from secondary UV photons in molecules photon$^{-1}$.}
    \tablefoottext{e}{Parameters used in the effective desorption yield parametrization.}
    \tablefoottext{f}{Chemical desorption factor for mono-product reactions.}
    \tablefoottext{g}{Chemical desorption factor for multi-product reactions.}
    \tablefoottext{h}{Strong uncertainties exist on the $\chi$ parameter as no clear relations between the diffusion energy and the desorption energy have been found \citep{2022ApJ...933L..16F}. It is usually between $0.3$ and $0.8$.}
    \tablefoottext{i}{Used for the definition number 2 of the reduced mass for diffusion (Table \ref{tab: switches}).}
    \tablefoottext{j}{Used for convergence criterion.}
    \tablefoottext{k}{Number of time steps at which DESTINY returns an output.}
    \tablefoottext{l}{Number of sigmas at which the BED is truncated.}
    }
    \label{tab: inputs}
\end{table}

\begin{table}[H]
    \centering
    \caption{Switches defined in DESTINY.}
    \begin{tabular}{l l}
    \hline
    \hline
    Physical switch & Default \\ \hline \hline
    is_ER_activated\tablefootmark{a} & True \\
    use_reac_tunneling\tablefootmark{b} & True \\
    use_diff_tunneling\tablefootmark{c} & True \\
    tunn_diff_reduced_mass_definition\tablefootmark{d} & 1 \\
    use_computed_f_chem_des\tablefootmark{e} & True \\
    use_diff_CR_heating\tablefootmark{f} & True \\
    use_reac_CR_heating\tablefootmark{g} & True \\
    is_surface_diff_to_des_ratio_species_specific\tablefootmark{h} & True \\
    use_computed_species_tf\tablefootmark{i} & True \\
    is_tf_species_specific\tablefootmark{j} & False \\
    \hline \hline
    Numerical switch & Default \\ \hline \hline
    Time_spacing\tablefootmark{k} & Log \\
    is_fixed_resolution\tablefootmark{l} & False \\
    
    \hline
    \end{tabular}
    \tablefoot{
    \tablefoottext{a}{Enables ER reactions.}
    \tablefoottext{b}{Enables reaction through tunneling.}
    \tablefoottext{c}{Enables diffusion through tunneling.}
    \tablefoottext{d}{Under configuration 1, the reduced mass of the system governing diffusion via quantum tunneling is set equal to the mass of the adsorbate, which corresponds to the physical assumption of an immobile substrate. Under configuration 2, the reduced mass is evaluated by accounting for both the adsorbate mass and the collective mass of $n_{H_2O}$ H$_2$O molecules.}
    \tablefoottext{e}{Utilizes the chemical desorption factor derived by \cite{2016A&A...585A..24M} specifically for mono-product reactive pathways, while the parameter $f_{\text{cd}}^{\text{multi}}$ remains implemented for multi-product channels. Conversely, if this switch is disabled, a uniform factor $f_{\text{cd}}$ is applied systematically across all reactive pathways, for which it is not specified in the chemical network.}
    \tablefoottext{f}{Enables diffusion through CR-heating.}
    \tablefoottext{g}{Enables reaction through CR-heating.}
    \tablefoottext{h}{Enables the choice of a specific $\chi$ parameter for each species.}
    \tablefoottext{i}{Uses computed BE-dependent trial frequency from the harmonic oscillator approximation (Eq. \ref{eq: trial_frequency}).}
    \tablefoottext{j}{Considered only if use_computed_species_tf is disabled. Enables the choice of a parametrized trial frequency specific for each species.}
    \tablefoottext{k}{"Log" for logarithmic or "Linear" for linear time spacing for the output time stepping.}
    \tablefoottext{l}{If true, utilizes a fixed number of bins, $n_{bins}$, for the BED discretizations. If false, utilizes a fixed BE resolution, $\Delta E$, for BED discretizations.}
    }
    \label{tab: switches}
\end{table}

The switches related to the definition of the trial frequency are worth to be commented. Given the current lack of constraints on the trial frequency of the adsorbates, the software architecture is designed to accommodate three distinct implementations for its parameterization: (i) a uniform, unique trial frequency applied across all chemical species, prescribed as a user-defined input parameter; (ii) species-specific trial frequencies, systematically assigned via independent user-defined inputs; or (iii) a fully discretized approach wherein the trial frequency is explicitly computed for each bin and for each individual species through Eq. \ref{eq: trial_frequency} based on the bin effective BE, as detailed in Sect.\,\ref{Sect: BED representation}.

It is worth mentioning that the definition of the trial frequency will be subject to further investigations in subsequent studies. The definition from Eq. \ref{eq: trial_frequency} based on the simple harmonic oscillator approximation should be compared to transition state theory-based definition, as suggested in \cite{2022ESC.....6..597M}.

\newpage

\section{Adaptive Gauss-Legendre quadrature scheme} \label{app: Gauss_Legendre}

To circumvent the underestimation problem for integrals lacking analytical closed forms, an adaptive Gauss-Legendre quadrature scheme has been applied. The roots of the associated Legendre polynomials lack closed-form algebraic expressions and are consequently resolved numerically via a Newton iterative algorithm. The quadrature scheme is subsequently executed with an increasing order $N$, which is sequentially doubled until the absolute difference between successive iterations falls below a predefined convergence tolerance, making the method adaptive. To optimize computational efficiency and avoid prohibitive runtime costs, the iteration is truncated if the quadrature order exceeds a maximum threshold value, at which point the last estimate is returned. Ultimately, the application of the complementary error function method and the Gauss-Legendre quadrature formalism to equations lacking exact closed-form solutions significantly enhances the numerical precision of the overall solver.

\section{Geometrical projection of the BE-resolved coverage state vector}\label{appendix: projection_formalism}

Physical admissibility of the BE-resolved coverages is enforced after each accepted CVODES time step through an Euclidean projection onto the intersection of a box-constrained domain, defined by $0 \le \theta_i(E_i) \le 1$ for every binding-energy bin, and a weighted simplex enforcing the global monolayer saturation constraint. This projection prevents numerical drift outside the physically admissible domain, thereby improving solver robustness and avoiding unphysical negative coverages or vacancy fractions of the form $1-\Theta$. To minimize computational overhead, an initial projection onto the box constraint is first performed, corresponding to a clipping operation of all coverages onto their admissible interval, and the weighted simplex constraint is subsequently evaluated. If the resulting state already satisfies the saturation constraint, no further correction is required. Otherwise, the projection consists in finding the admissible state vector
$\mathbf{x}$ closest to the unconstrained CVODES solution
$\mathbf{y}$. The projection is defined as

\begin{equation}
\min_{\mathbf{x}}
\frac12 \|\mathbf{x}-\mathbf{y}\|_2^2
\end{equation}

subject to

\begin{equation}
0 \le x_j \le 1,
\qquad
\sum_j w_j x_j \le 1.
\end{equation}

Since the objective function is strictly convex and the feasible set is convex, the projected solution is unique. The problem is formulated within a Lagrangian framework through

\begin{align}
    \mathcal{L}(\mathbf{x},\lambda,\boldsymbol{\alpha},\boldsymbol{\beta})
    &=
    \frac12 \sum_j (x_j-y_j)^2
    +
    \lambda
    \left(
    \sum_j w_j x_j -1
    \right) \\\notag
    &-
    \sum_j \alpha_j x_j
    +
    \sum_j \beta_j (x_j-1),
\end{align}

where $\lambda$ denotes the Lagrange multiplier associated with the weighted saturation constraint, while $\alpha_j$ and $\beta_j$ enforce the lower and upper box bounds, respectively. The corresponding KKT conditions yield the interior solution

\begin{equation}
    x_j = y_j - \lambda w_j,
\end{equation}

which remains valid as long as all projected components remain strictly inside the box bounds. In this case, the multiplier admits the analytical expression

\begin{equation}
    \lambda
    =
    \frac{\sum_j w_j y_j -1}
         {\sum_j w_j^2}.
\end{equation}

When one or more components reach an active box constraint, the interior solution is no longer valid and the projected state becomes

\begin{equation}
    x_j(\lambda)
    =
    \mathrm{clip}
    \left(
    y_j-\lambda w_j,
    \,0,\,
    1
    \right).
\end{equation}

Substituting this expression into the weighted simplex constraint defines the monotonic scalar function

\begin{equation}
    \phi(\lambda)
    =
    \sum_j
    w_j
    \,
    \mathrm{clip}
    \left(
    y_j-\lambda w_j,
    \,0,\,
    1
    \right),
\end{equation}

and the required multiplier is obtained as the unique root of

\begin{equation}
    \phi(\lambda)-1=0.
\end{equation}

The monotonicity of $\phi(\lambda)$ establishes a bijection between the value of the Lagrange multiplier and the weighted saturation of the projected state, reducing the multidimensional projection problem to the determination of a single scalar unknown. The latter is obtained through a bracketing-and-bisection procedure, successively shrinking the interval $[\lambda_L,\lambda_U]$ until convergence, while returning the upper bracket $\lambda_U$ to guarantee satisfaction of the saturation constraint to machine precision.

\newpage

\section{Reduced solid-phase chemical network used in this paper}\label{app: benchmark_network}

Table \ref{tab: BEDs} gives the BED parameters for the 26 solid-phase species considered in this study. We note that the sticking coefficient of all included solid-phase species are set to 1, except for H and H$_2$ where the temperature-dependent values are used \citep{2010JChPh.133j4507M, 2012A&A...538A.128C}, as performed in Nautilus \citep{2021A&A...652A..63W}. The related reduced-reaction network is given in table \ref{tab: network}. 

\begin{table}[H]
\caption{\label{tab: BEDs}Reduced set of surface species and BEDs parameters used in this work: $\mu_{\rm BE}$ column give the mean BE, while $\sigma_{\rm BE}$ stands for the BED standard deviation.}
\centering
\begin{tabular}{l l l l}
\\
\hline
Species & Distribution & $\mu_{\rm BE}$ (K) & $\sigma_{\rm BE}$ (K) \\
\hline
\hline
H$_2$      & Gaussian & 334  & 67   \\
H          & Gaussian & 650  & 130  \\
O          & Gaussian & 1600 & 320  \\
OH         & Gaussian & 4600 & 920  \\
H$_2$O     & Gaussian & 5600 & 1120 \\
N          & Gaussian & 720  & 142  \\
NH         & Gaussian & 2600 & 520  \\
NH$_2$     & Gaussian & 3200 & 640  \\
NH$_3$     & Gaussian & 5500 & 1100 \\
     & GMM2 & -- & -- \\
NO         & Gaussian & 1600 & 320  \\
C          & Gaussian & 10.000 & 1500  \\
CN         & Gaussian & 2800 & 540  \\
HCN        & Gaussian & 3700 & 740  \\
CO         & Gaussian & 1300 & 260  \\
      & Gaussian & 1407 & 339  \\
CH$_4$     & Gaussian & 960  & 192  \\
     & Gaussian & 1388  & 242  \\
CH$_3$     & Gaussian & 1600 & 320  \\
CH$_2$     & Gaussian & 1400 & 280  \\
CH         & Gaussian & 925  & 185  \\
O$_2$      & Gaussian & 1200 & 240  \\
HCO        & Gaussian & 2400 & 480  \\
H$_2$CO    & Gaussian & 4500 & 900  \\
CO$_2$     & Gaussian & 2600 & 520  \\
HOCO       & Gaussian & 2000 & 400  \\
HCOOH      & Gaussian & 5570 & 1114 \\
CH$_3$O    & Gaussian & 4400 & 880  \\
CH$_2$OH   & Gaussian & 4400 & 880  \\
CH$_3$OH   & Gaussian & 5000 & 1000  \\
\hline
\end{tabular}
\tablefoot{
For H$_2$, the H$_2$--H$_2$ encounter binding energy is set to 23 K and the corresponding diffusion-to-desorption ratio $\chi$ is set to 0.4; both quantities are user-customizable parameters. NH$_3$ can alternatively be described by a Gaussian mixture model (GMM2), with component means at 2803 ($\sigma_{\rm BE, \, k=1}$ 613 K), 5620 K ($\sigma_{\rm BE, \, k=2}$ 928 K) and corresponding weights of 0.16 and 0.84, respectively, based on results in \cite{2025A&A...698A.284G}. All BE means come from single values reported in KIDA database\footnote{\url{https://kida.astrochem-tools.org}}, except for the second, alternative values for CO, CH$_4$ and NH$_3$, taken from \cite{2025A&A...698A.284G}. Selected KIDA values are either taken from \cite{2017MolAs...6...22W} when available, or from the original OSU gas-grain code from Eric Herbst group in 2006 \citep{2006A&A...457..927G}. KIDA values are considered for the benchmark against Nautilus code (Sect. \ref{subsec: result_comp_to_Naut}), while alternative values for CO, CH$_4$ and NH$_3$ are considered for the assessment of the BED discretization effect (Sect. \ref{subsec: result_discretiz}). Standard deviation values are taken to be of 20\% of the respective means, except for C (15\%) as well as for alternative values taken from from \cite{2025A&A...698A.284G}.  
Formation enthalpies used for the computation of chemical desorption fraction (Eq. \ref{eq: f_cd}) are directly taken from KIDA and are not explicitely listed here.}
\end{table}

\begin{table}[H]
\caption{\label{tab: network}Reduced benchmark surface reaction network used in this work.}
\centering
\begin{tabular}{lll}\\
\hline
Reaction & $E_{\rm A}$ (K) & $f_{\rm cd}$ \\
\hline
\hline

H + H $\rightarrow$ H$_2$                             & 0    & Eq.~\ref{eq: f_cd} \\
H + O $\rightarrow$ OH                               & 0    & 0.30 \\
H + OH $\rightarrow$ H$_2$O                          & 0    & 0.25 \\
H$_2$ + OH $\rightarrow$ H$_2$O + H                 & 2100 & Eq.~\ref{eq: f_cd} \\
N + O $\rightarrow$ NO                               & 0    & Eq.~\ref{eq: f_cd} \\
N + C $\rightarrow$ CN                               & 0    & Eq.~\ref{eq: f_cd} \\
H + CN $\rightarrow$ HCN                             & 0    & Eq.~\ref{eq: f_cd} \\
N + H $\rightarrow$ NH                               & 0    & Eq.~\ref{eq: f_cd} \\
NH + H $\rightarrow$ NH$_2$                          & 0    & Eq.~\ref{eq: f_cd} \\
NH$_2$ + H$_2$ $\rightarrow$ NH$_3$ + H            & 6300 & Eq.~\ref{eq: f_cd} \\
NH$_2$ + H $\rightarrow$ NH$_3$                      & 0    & Eq.~\ref{eq: f_cd} \\
C + O $\rightarrow$ CO                               & 0    & Eq.~\ref{eq: f_cd} \\
C + OH $\rightarrow$ CO + H                          & 0    & Eq.~\ref{eq: f_cd} \\
C + O$_2$ $\rightarrow$ CO + O                       & 0    & Eq.~\ref{eq: f_cd} \\
O + HCO $\rightarrow$ CO + OH                        & 0    & Eq.~\ref{eq: f_cd} \\
O + HCO $\rightarrow$ CO$_2$ + H                     & 0    & Eq.~\ref{eq: f_cd} \\
OH + HCO $\rightarrow$ HCOOH                         & 0    & Eq.~\ref{eq: f_cd} \\
H + HOCO $\rightarrow$ HCOOH                         & 0    & Eq.~\ref{eq: f_cd} \\
H + HOCO $\rightarrow$ H$_2$ + CO$_2$              & 0    & Eq.~\ref{eq: f_cd} \\
H + HOCO $\rightarrow$ CO + H$_2$O                 & 0    & Eq.~\ref{eq: f_cd} \\
H + CO $\rightarrow$ HCO                             & 2500 & Eq.~\ref{eq: f_cd} \\
O + CO $\rightarrow$ CO$_2$                          & 1000 & Eq.~\ref{eq: f_cd} \\
OH + CO $\rightarrow$ HOCO                           & 150  & Eq.~\ref{eq: f_cd} \\
OH + CO $\rightarrow$ CO$_2$ + H                     & 150  & Eq.~\ref{eq: f_cd} \\
H + H$_2$CO $\rightarrow$ CH$_2$OH                  & 5400 & Eq.~\ref{eq: f_cd} \\
H + H$_2$CO $\rightarrow$ CH$_3$O                   & 2200 & Eq.~\ref{eq: f_cd} \\
H + H$_2$CO $\rightarrow$ H$_2$ + HCO              & 1740 & Eq.~\ref{eq: f_cd} \\
OH + H$_2$CO $\rightarrow$ H$_2$O + HCO            & 0    & Eq.~\ref{eq: f_cd} \\
O + CH $\rightarrow$ HCO                             & 0    & Eq.~\ref{eq: f_cd} \\
H + HCO $\rightarrow$ H$_2$CO                        & 0    & Eq.~\ref{eq: f_cd} \\
O$_2$ + CH $\rightarrow$ HCO + O                     & 0    & Eq.~\ref{eq: f_cd} \\
O + CH$_2$ $\rightarrow$ H$_2$CO                    & 0    & Eq.~\ref{eq: f_cd} \\
O$_2$ + CH$_2$ $\rightarrow$ H$_2$CO + O           & 0    & Eq.~\ref{eq: f_cd} \\
C + H $\rightarrow$ CH                               & 0    & Eq.~\ref{eq: f_cd} \\
C + H$_2$ $\rightarrow$ CH$_2$                       & 0    & Eq.~\ref{eq: f_cd} \\
CH + H $\rightarrow$ CH$_2$                          & 0    & Eq.~\ref{eq: f_cd} \\
CH$_2$ + H $\rightarrow$ CH$_3$                      & 0    & Eq.~\ref{eq: f_cd} \\
CH$_2$ + H$_2$ $\rightarrow$ CH$_3$ + H            & 3530 & Eq.~\ref{eq: f_cd} \\
CH$_3$ + H $\rightarrow$ CH$_4$                      & 0    & Eq.~\ref{eq: f_cd} \\
CH$_4$ + H $\rightarrow$ CH$_3$ + H$_2$            & 5940 & Eq.~\ref{eq: f_cd} \\
CH$_3$ + H$_2$ $\rightarrow$ CH$_4$ + H            & 6440 & Eq.~\ref{eq: f_cd} \\
OH + CH$_3$ $\rightarrow$ CH$_3$OH                  & 0    & Eq.~\ref{eq: f_cd} \\
H + CH$_3$O $\rightarrow$ CH$_3$OH                   & 0    & Eq.~\ref{eq: f_cd} \\
H + CH$_2$OH $\rightarrow$ CH$_3$OH                  & 0    & Eq.~\ref{eq: f_cd} \\

\hline
\end{tabular}
\tablefoot{All activation energies come from KIDA database}
\end{table}

\section{Insights for DESTINY benchmark against Nautilus}\label{app: Naut_comp}

\begin{figure}
    \centering
    \includegraphics[width=0.9  \linewidth]{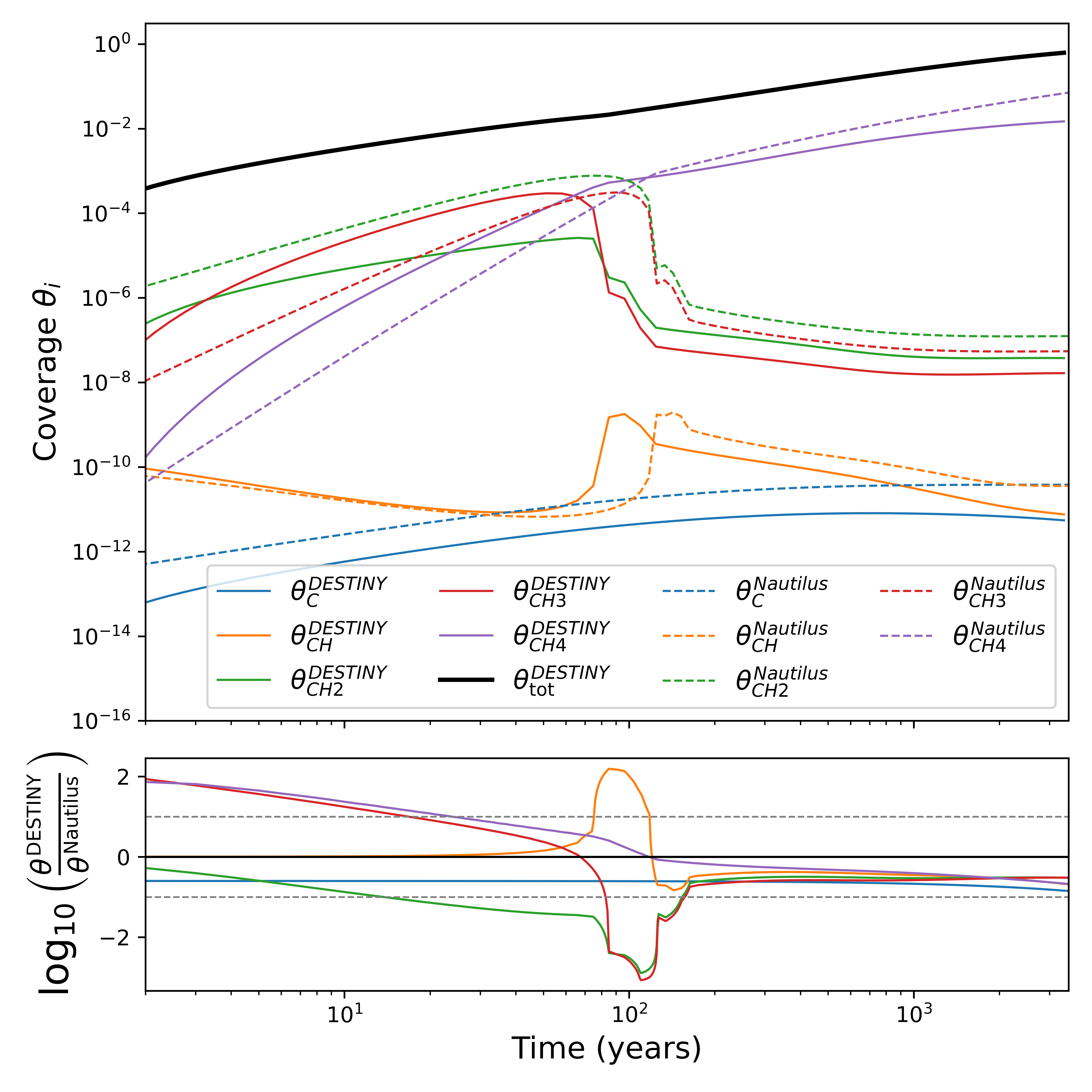}
    \caption{Comparison between the surface abundances predicted by DESTINY and Nautilus in the single-BE limit for the CH$_x$ chain. }
    \label{fig: compa_to_Naut_CHx_chain}
\end{figure}

As discussed in the main text, CH$_4$ exhibits the largest deviation from Nautilus, particularly at early times when the monolayer correction $(1 - \Theta)$ has negligible influence on the accretion-limited chemistry. Fig. \ref{fig: compa_to_Naut_CHx_chain} compares DESTINY and Nautilus coverages for atomic C and the successive CH$_x$ hydrogenated products. CH is the least affected species, displaying noticeable differences only during its transient coverage rise. Within the reduced network, it is also the only CH$_x$ member whose coverage is not directly influenced by H$_2$-driven chemistry.

Relative to Nautilus, DESTINY predicts systematically lower coverages for both C and CH$_2$, the latter being the direct product of the barrierless reaction C + H$_2 \rightarrow$ CH$_2$. The opposite behavior is observed for CH$_3$ and CH$_4$, which are linked to CH$_2$ through the reactions CH$_2$ + H$_2 \rightarrow$ CH$_3$ + H ($E_a = 3530$ K), CH$_3$ + H$_2 \rightarrow $ CH$_4$ + H ($E_a = 6400$ K), and the barrierless additions CH$_2$ + H $\rightarrow$ CH$_3$ and CH$_3$ + H $\rightarrow$ CH$_4$. This pattern indicates enhanced consumption of C and CH$_2$ in favor of CH$_3$ and CH$_4$ formation. Although CH$_3$O is directly produced from CH$_3$ through barrierless O addition, its abundance remains in good agreement with Nautilus. This reflects the limited availability of atomic O, which is efficiently converted into OH and subsequently H$_2$O.

\begin{figure}
    \centering
    \includegraphics[width=1.0  \linewidth]{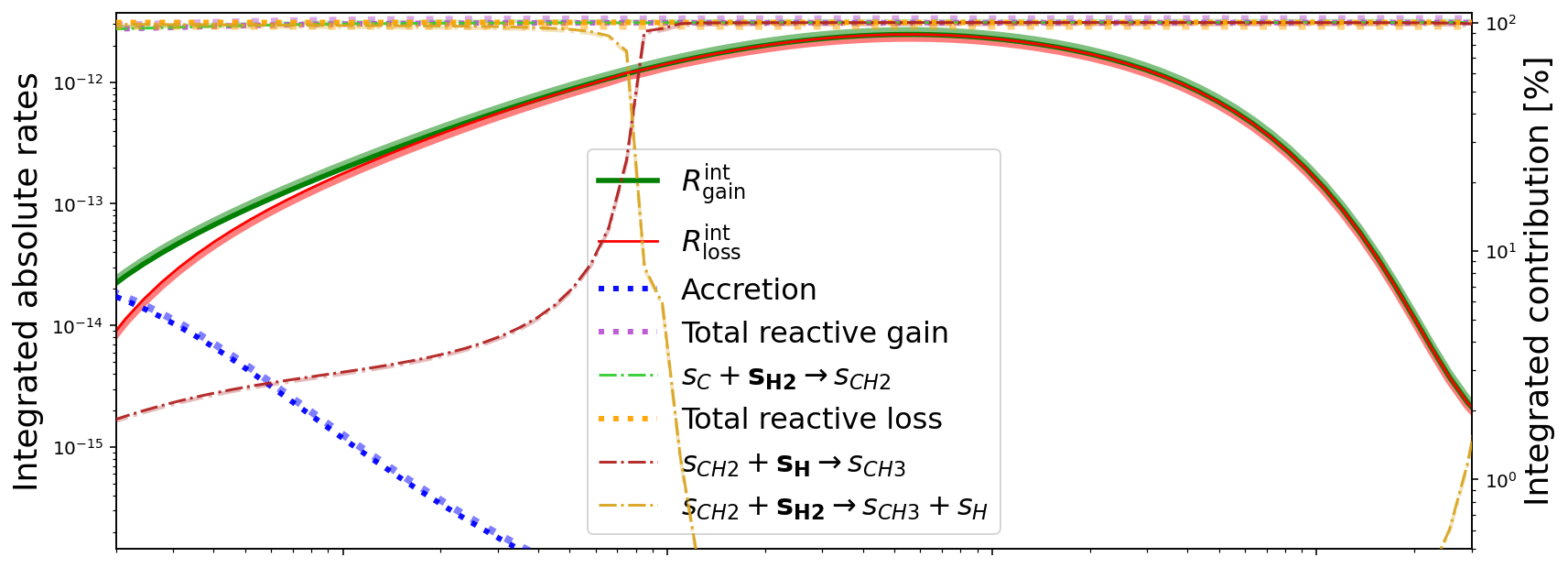}
    \caption{Absolute gain and loss rate for solid-phase CH$_2$ (left y axis), and related relative contributions (right y axis). Gain contribution are given in warm colors, while loss contributions are related to the cold color lines. Highlighted species stands for the attacking reactant.}
    \label{fig: CH2_routes}
\end{figure}

Figure \ref{fig: CH2_routes} shows the absolute gain and loss rates of $\Theta_{CH_2}$, along with its dominant formation and destruction channels. It reveals the central role of H$_2$-driven chemistry in the CH$_x$ chain, especially at early times when the monolayer correction $(1-\Theta)$ has a negligible impact on the accretion-limited chemistry.

\begin{figure}
    \centering
    \includegraphics[width=0.9  \linewidth]{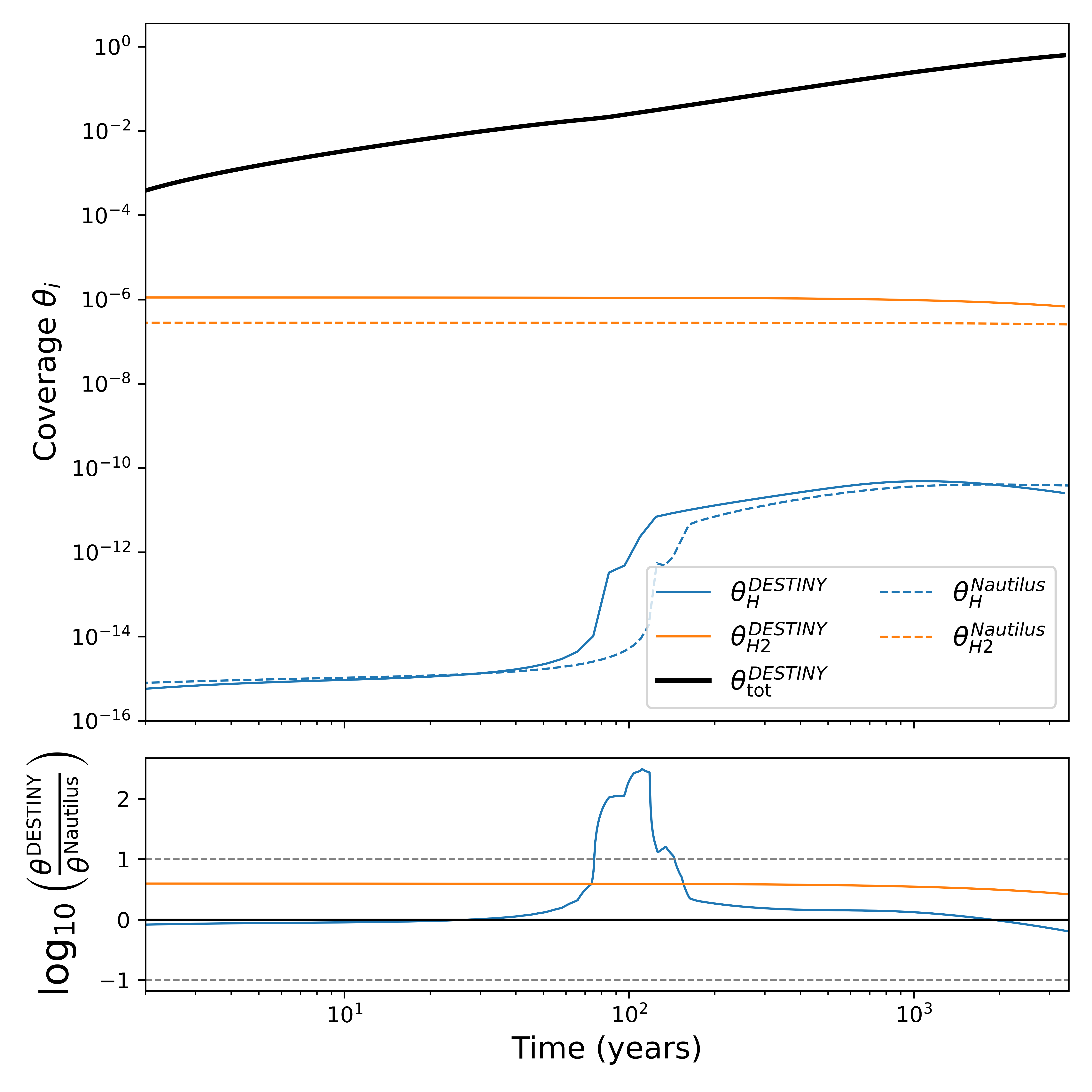}
    \caption{Comparison between the surface abundances predicted by DESTINY and Nautilus in the single-BE limit for H and H$_2$. }
    \label{fig: compa_to_Naut_H_H2}
\end{figure}

Finally, Fig. \ref{fig: compa_to_Naut_H_H2} compared the H and H$_2$ coverages predicted by Nautilus and DESTINY, still within the single-bin limit. The systematically larger H$_2$ abundance obtained with DESTINY is arguably explained by the H$_2$ encounter effect and the inherent dynamical coupling between H$_2$ desorption and diffusion rates. In Nautilus, prescriptions from Eqs. 1 and 2 in \citet{Hincelin2015} are considered; H$_2$--H$_2$ encounter rate $k_{H_2--H_2}$, describing the rate of two H$_2$ diffusing on the same lattice site ($k_{H_2} + k_{H_2}$), acts as an effective access rate to an encounter-desorption loss channel. Meanwhile, the desorption efficiency is evaluated through a competition including non-thermal desorption channels, as prescribed by Eq. 2 in \citet{Hincelin2015}. In our scheme, non-thermal desorption are not concerned by the TFC competition. Indeed, so-called non-thermal desorption processes, namely here photodesorption and CR-sputtering, (i) do not feature any BE dependancy of their rates, at least within the current parameterization and underlying hypotheses (Appendix \ref{app: ice_proc}), and (ii) depends on external sources, analogously to the ER mechanism. These are therefore excluded from the TFC. The non-thermal rates are additionally enhanced by a factor $10^{30}$ in Nautilus source code. As a result, the H$_2$ encounter effect in Nautilus and other implementations based on prescriptions in \citet{Hincelin2015} are (i) non branched towards an enhanced H$_2$ mobility, with an un-perturbed $k_{H_2--H_2}$ encounter-rate coefficient, and (ii) more efficiently converted into desorption. This combination leads to a slower encounter-driven surface exploration and a more efficient depletion of surface H$_2$. In DESTINY, by contrast, each H$_2$--H$_2$ encounter is treated as a double-filled configuration in which the reduced encounter BE affects both desorption and re-diffusion. A fraction of the encounter flux is therefore redirected toward enhanced diffusion rather than immediate loss, allowing H$_2$ to resume surface exploration and potentially trigger further encounters. This self-consistent coupling can reduce the net efficiency of the H$_2$ encounter effect on the H$_2$ desorption, and explain the higher retained $\Theta_{H_2}$.

A transient enhancement of the H surface coverage is also observed in the DESTINY results around 100 years. This increase coincides with the sharp temporary deviations exhibited by all hydrogenated species, including NH$_3$ discussed in the main text. Given the transient nature of this feature, it will not be examined further in the present work.

\section{Insights onto the comparison between single-BE limit versus 10 bins coverages} \label{app: multi_bin}

\begin{figure}[H]
    \centering
    \includegraphics[width=0.9\linewidth]{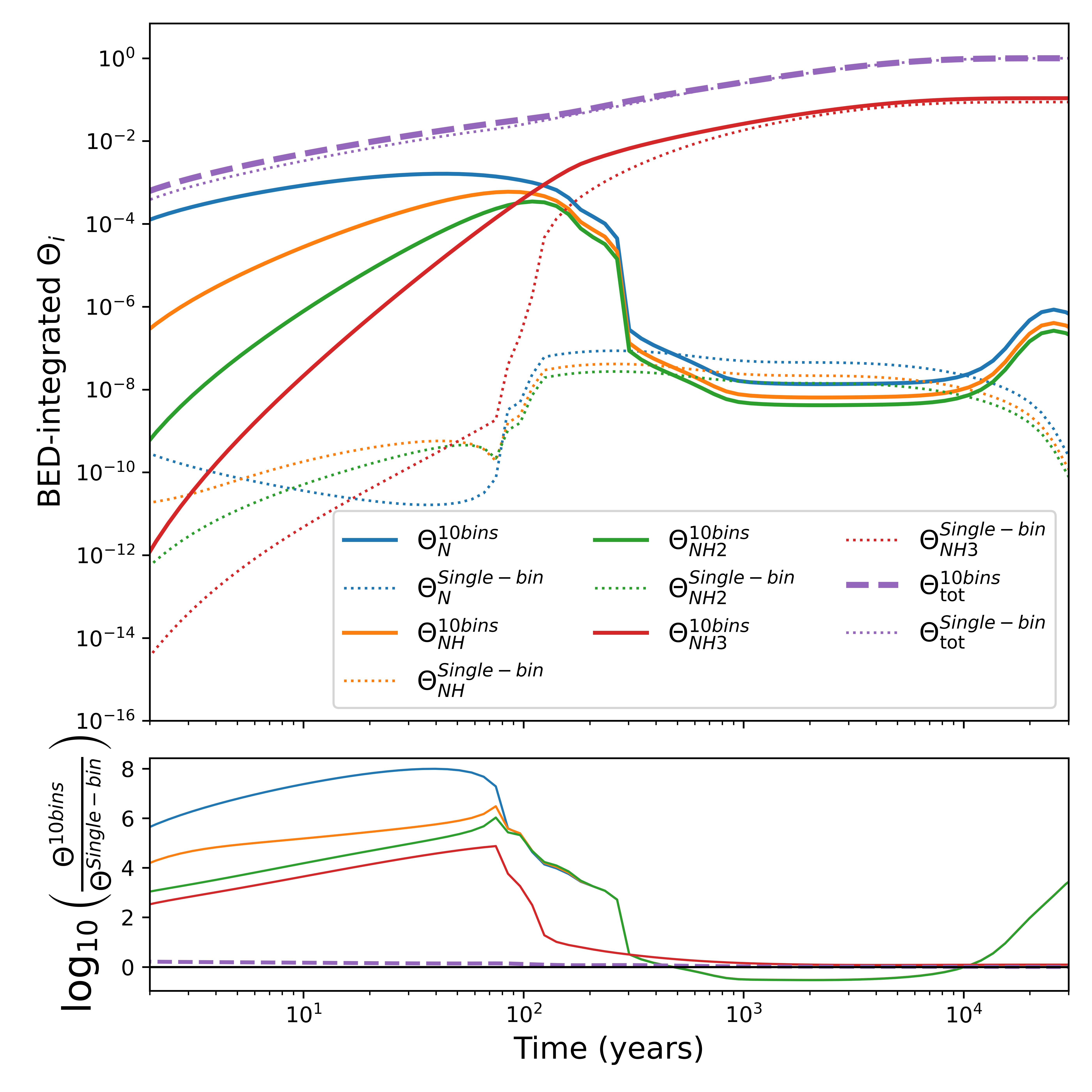}
    \caption{Comparison between the surface abundances predicted by DESTINY for the single-bin and 10-bins simulations for the NH$_x$ chain.}
    \label{fig: NH3_discretiz}
\end{figure}

This appendix provides complementary resources regarding the divergence between the single-bin and multi-bin configurations discussed in Sect. \ref{subsec: result_discretiz} for the selected key chemical species. The trapping mechanism invoked to explain the enhanced synthesis of NH$_3$ molecules is computationally demonstrated by a corresponding decrease in its integrated diffusion contribution, alongside an increase in the surface abundances of each intermediate species within its hydrogenation chain, as illustrated in Fig. \ref{fig: NH3_discretiz}.

\begin{figure}[H]
    \centering
    \includegraphics[width=1.0\linewidth]{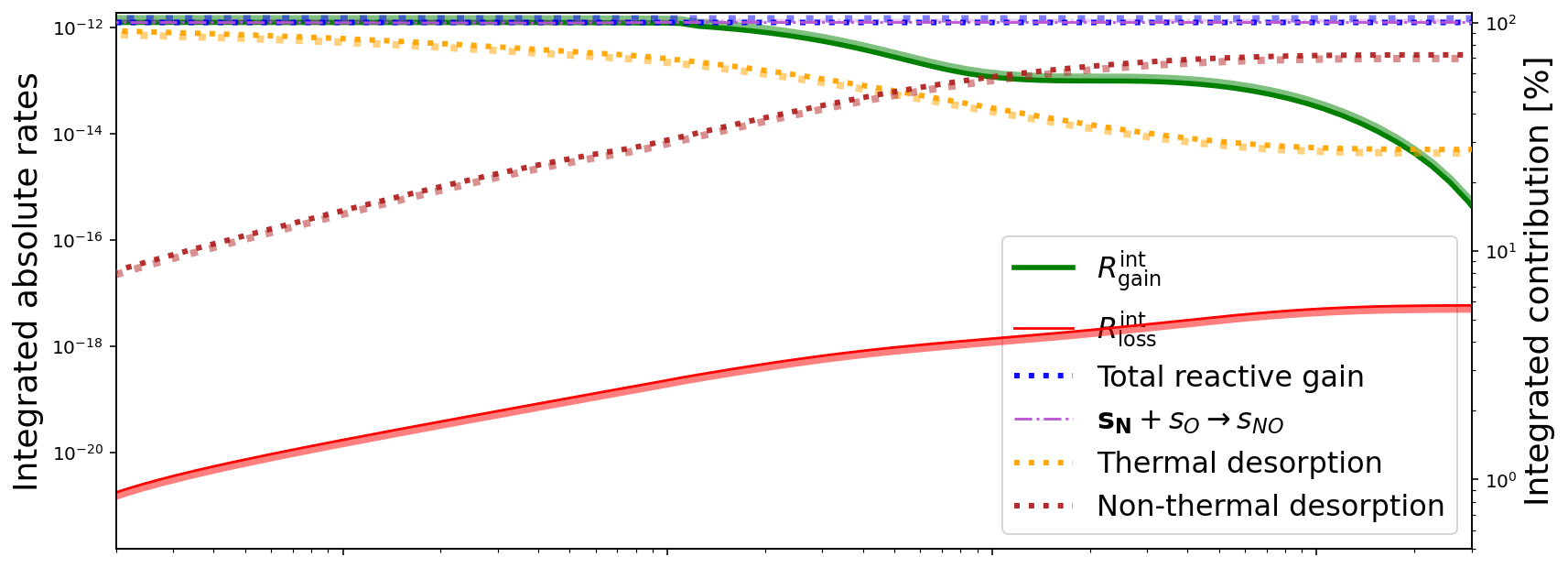}
    \caption{Absolute gain and loss rate for solid-phase NO (left y axis) in the single-bin model, together with the related relative contributions (scale on the right y axis).}
    \label{fig: NO_1bin}
\end{figure}

\begin{figure}[H]
    \centering
    \includegraphics[width=1.0\linewidth]{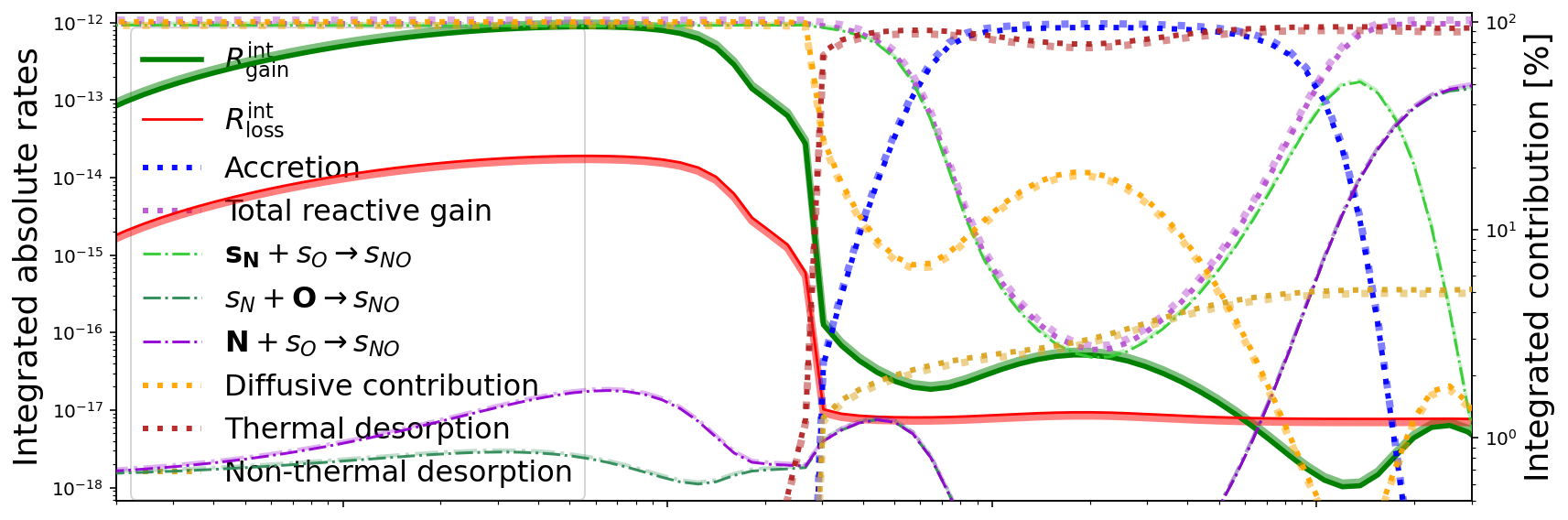}
    \caption{Absolute gain and loss rate for solid-phase NO (left y axis) in the 10-bins model, together with the related relative contributions (scale on the right y axis).}
    \label{fig: NO_10bins}
\end{figure}

The slight shift in favor of the ER mechanism, as well as the amplified thermal desorption for NO species, is directly inferred from the integrated contribution of each microphysical process and the integrated absolute rates of the single- and multi-bin simulations, as shown in Fig. \ref{fig: NO_1bin} - \ref{fig: NO_10bins}.

\begin{figure}[H]
    \centering
    \includegraphics[width=0.9\linewidth]{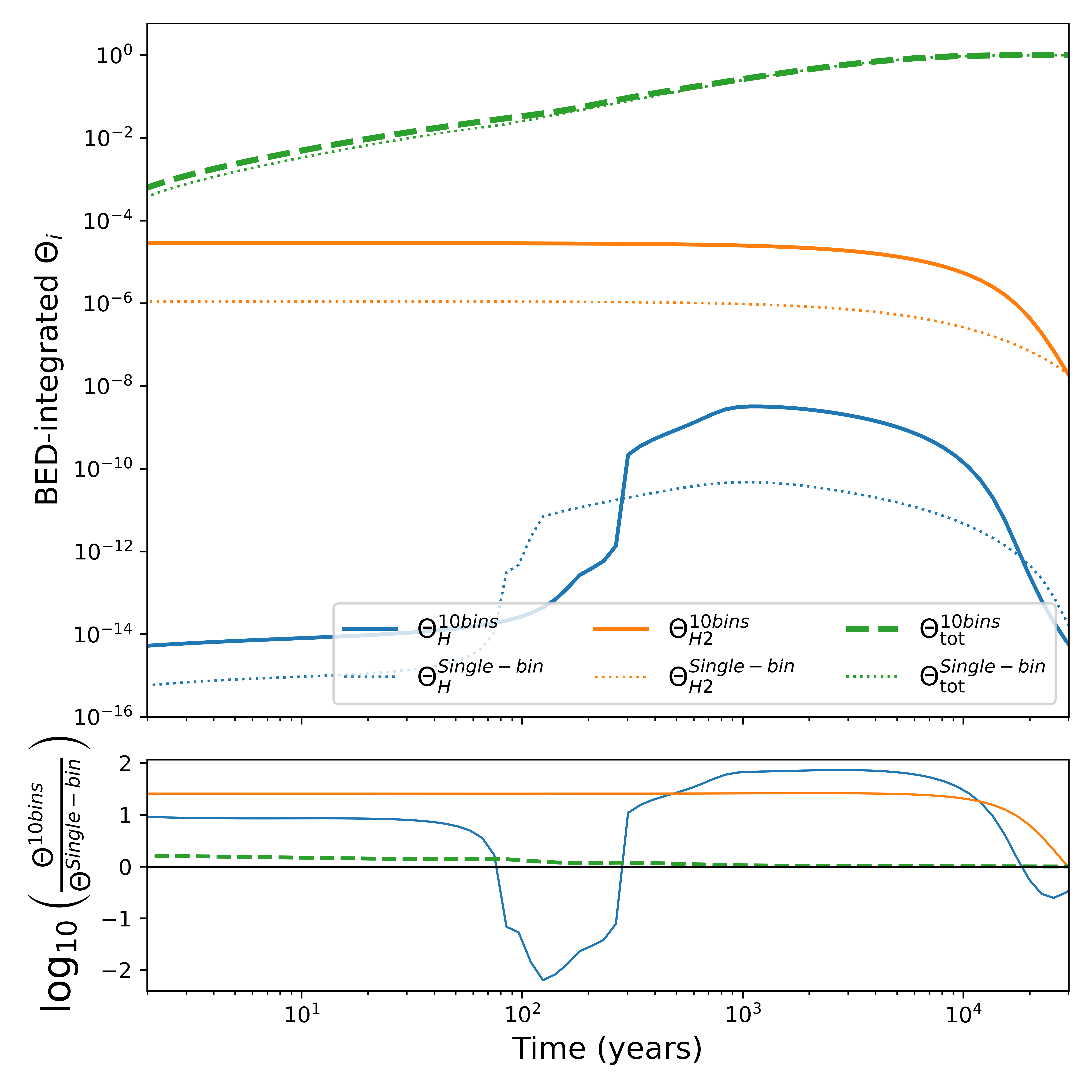}
    \caption{Comparison between the surface abundances predicted by DESTINY with 10 bins as compared to the single-BE limit for H and H$_2$.}
    \label{fig: H_multibin}
\end{figure}

The overabundance of atomic hydrogen in the first period and its slight underabundance around 100 years is illustrated in \ref{fig: H_multibin}.

\begin{figure}[H]
    \centering
    \includegraphics[width=1.0\linewidth]{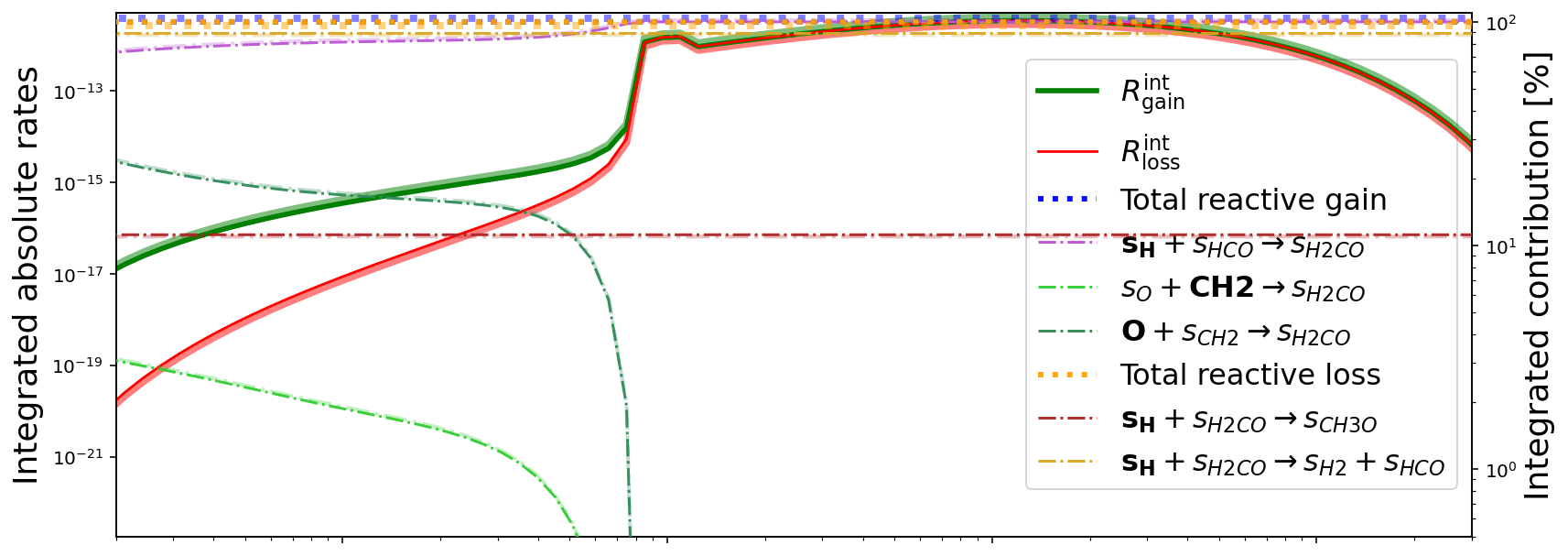}
    \caption{(Top panel) Absolute gain and loss rate for solid-phase H$_2$CO (left y axis) in the single-bin model, together with the related relative contributions (scale on the right y axis). (Bottom panel) Gain to loss balance.}
    \label{fig: H2CO_1bins}
\end{figure}

\begin{figure}[H]
    \centering
    \includegraphics[width=1.0\linewidth]{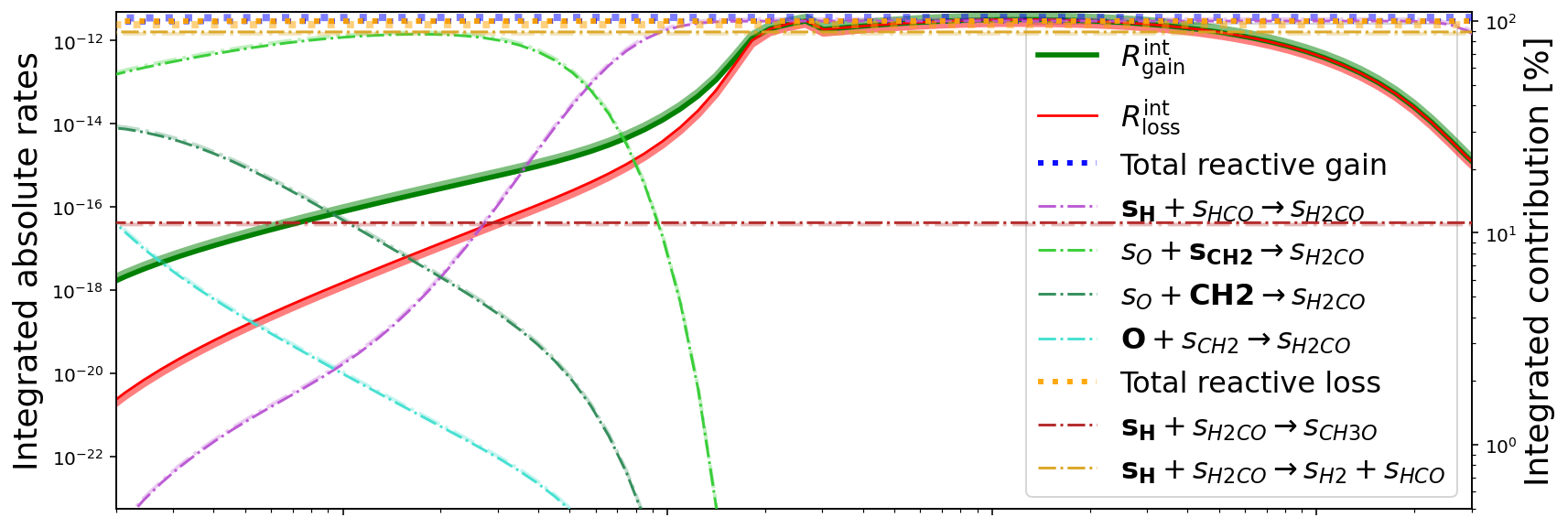}
    \caption{(Top panel) Absolute gain and loss rate for solid-phase H$_2$CO (left y axis) in the 10-bins model, together with the related relative contributions (scale on the right y axis). (Bottom panel) Gain to loss balance.}
    \label{fig: H2CO_10bins}
\end{figure}

The switch in the chemical pathways in favor of the CH$_2$ + O $\rightarrow$ H$_2$CO reaction for the H$_2$CO between the single-BE limit and the multi-bin configuration is shown in Fig. \ref{fig: H2CO_1bins} – Fig. \ref{fig: H2CO_10bins}.

\end{appendix}

\end{document}